\documentclass[twocolumn]{aastex62}
\usepackage{times}
\usepackage{amsmath}
\usepackage{graphicx}
\usepackage{subfigure}
\usepackage{placeins}
\usepackage{hyperref}
\usepackage{gensymb}
\usepackage{upgreek}
\usepackage{multirow}
\usepackage[normalem]{ulem}
\newcommand{\kms}{km~s$^{\mathrm{-1}}$}

\newcommand{\HI}{H{\sevenrm\,I}}

\newcommand{\RHI}{$R_{\rm H\, \textsc{i}}$}

\newcommand{\MHI}{$M_{\rm H\, \textsc{i}}$}
\newcommand{\mdyn}{$M_{\rm dyn}$}
\newcommand{\mtwo}{$M_{200}$}
\newcommand{\msun}{$M_{\mathrm{\odot}}$}

\newcommand{\vrot}{$V_{\rm rot}$}

\newcommand{\mstar}{$M_{*}$}

 \font\sevenrm=cmr7 scaled 1000

\newcommand{\stkout}[1]{\ifmmode\text{\sout{\ensuremath{#1}}}\else\sout{#1}\fi}

\newdimen\digitwidth    
\setbox1=\hbox{0}       
\digitwidth=\wd1        
\catcode`"=\active      
\def"{\kern\digitwidth}

\begin{document}

\title{On the Determination of Rotation Velocity and Dynamical Mass of Galaxies Based on Integrated \HI\ Spectra}

\shorttitle{Galaxy \HI\ Spectra}
\shortauthors{YU, HO \& WANG}

\author{Niankun Yu}
\affiliation{Kavli Institute for Astronomy and Astrophysics, Peking University, Beijing 100871, China}
\affiliation{Department of Astronomy, School of Physics, Peking University, Beijing 100871, China}

\author{Luis C. Ho}
\affiliation{Kavli Institute for Astronomy and Astrophysics, Peking University, Beijing 100871, China}
\affiliation{Department of Astronomy, School of Physics, Peking University, Beijing 100871, China}

\author{Jing Wang}
\affiliation{Kavli Institute for Astronomy and Astrophysics, Peking University, Beijing 100871, China}
\affiliation{Department of Astronomy, School of Physics, Peking University, Beijing 100871, China}
\email{Email: niankun@pku.edu.cn}

\begin{abstract}
The integrated 21~cm \HI\ emission profile of a galaxy encodes valuable information on the kinematics, spatial distribution, and dynamical state of its cold interstellar medium.  The line width, in particular, reflects the rotation velocity of the galaxy, which, in combination with a size scale, can be used to constrain the dynamical mass of the system.  We introduce a new method based on the concept of the curve-of-growth to derive a set of robust parameters to characterize the line width, asymmetry, and concentration of the integrated \HI\ spectra.  We use mock spectra to evaluate the performance of our method, to estimate realistic systematic uncertainties for the proposed parameters, and to correct the line widths for the effects of instrumental resolution and turbulence broadening.  Using a large sample of nearby galaxies with available spatially resolved kinematics, we demonstrate that the newly defined line widths can predict the rotational velocities of galaxies to within an accuracy of $\lesssim 30$ \kms.  We use the calibrated line widths, in conjunction with the empirical relation between the size and mass of \HI\ disks, to formulate a prescription for estimating the dynamical mass within the \HI-emitting region of gas-rich galaxies.  Our formalism yields dynamical masses accurate to $\sim 0.3$ dex based solely on quantities that can be derived efficiently and robustly from current and future extragalactic \HI\ surveys.  We further extend the dynamical mass calibration to the scale of the dark matter halo.  
\end{abstract}


\section{Introduction}\label{sec:intro}

Neutral atomic hydrogen (\HI), as the raw material that ultimately forms stars and builds central black holes, plays an extremely important role in galaxy formation and evolution (e.g., \citealt{Krumholz2012ApJ...759....9K, Hess2018AAS...23121005H, Pokhrel2019AAS...23346005P}).  During the past two decades, single-dish surveys such as HIPASS (\citealt{Koribalski2004}; \citealt{HICAT}; \citealt{Wong2006MNRAS.371.1855W}), GASS \citep{Catinella2008AIPC.1035..252C}, and ALFALFA (\citealt{Haynes+2011AJ....142..170H,Haynes2018ApJ...861...49H}) have accumulated integrated \HI\ spectra for more than $10^4$ galaxies, and even larger samples are anticipated from upcoming surveys using ASKAP \citep{Norris2010AAS...21560405N}, Apertif \citep{Adams2018AAS...23135404A}, and FAST \citep{Nan2011IJMPD..20..989N}.

Apart from providing basic measures of radial velocity and hydrogen gas content, the integrated \HI\ 21~cm emission line profile encodes rich information pertaining to the internal kinematics of the galaxy, the spatial distribution of the gas, its dynamical state, and even rough indications of its external environment.  If the gas is in dynamic equilibrium, the line width traces the rotation velocity of the gas, which, in combination with an assumed or estimated size, can be used to calculate the galaxy's dynamical mass (e.g., \citealt{Casertano1980A&A....81..371C}; \citealt{BroeilsRhee1997}; \citealt{Ho2008ApJS..177..103H}; \citealt{Garcia-Appadoo2009MNRAS.394..340G}; \citealt{Koribalski2018MNRAS.478.1611K}).  This is obviously very attractive, in view of the economy it affords compared to securing spatially resolved velocity fields of the gas for detailed dynamical modeling (e.g., \citealt{deBlok2008AJ....136.2648D}; \citealt{Oh2011AJ....141..193O}; \citealt{Li2019MNRAS.482.5106L}), and it is complementary to mass estimates based on weak lensing analysis (e.g., \citealt{Mandelbaum2006MNRAS.368..715M}) and other kinematic tracers (e.g., \citealt{More2011MNRAS.410..210M, vanDokkum2018Natur.555..629V}).  Since the seminal work of \citet{TullyFisher1977AA....54..661T}, the correlation between galaxy rotation velocity and its stellar luminosity has evolved from its original intent as a distance indicator to an ever-more refined probe of the properties of dark matter halos and other aspects of galaxy formation physics (e.g., \citealt{McGaugh2000ApJ...533L..99M}; \citealt{Bell2001ApJ...550..212B}; \citealt{Freeman2004PASA...21..382F}; \citealt{Gnedin2007ApJ...671.1115G}; \citealt{Blanton2008ApJ...682..861B}; \citealt{Vogelsberger2014MNRAS.444.1518V}; \citealt{Schaye2015MNRAS.446..521S}; \citealt{Dutton2017MNRAS.467.4937D}).  However, opinion varies as to how the line width should be measured (e.g., \citealt{Matthews1998AJ....116.1169M}; \citealt{Koribalski2004, Courtois+2009AJ....138.1938C, Courtois2011MNRAS.414.2005C}), how it relates to the rotation velocity of the galaxy (e.g., \citealt{Rhee1996PhDT........36R}; \citealt{Meyer2008MNRAS.391.1712M}), and how different definitions of the line width can cause systematic differences in slope, intercept, and scatter of the derived Tully-Fisher relation (e.g., \citealt{McGaugh2012AJ....143...40M}; \citealt{Bradford2016ApJ...832...11B}).

Even in the absence of any direct spatial information, the very shape of the \HI\ spectrum provides clues regarding the spatial distribution of the gas in a galaxy. While many \HI\ profiles have the classic shape of a double-horn, galaxies exhibit a much wider range of profiles.  What can be learned from this diversity of line shapes? The \HI\ disk, highly extended and fragile, can be easily distorted by external perturbations such as tidal interactions (\citealt{Hess2017MNRAS.464..957H}; \citealt{Sorgho2017MNRAS.464..530S}; \citealt{Bok2019MNRAS.484..582B}), ram pressure stripping (\citealt{Scott2010MNRAS.403.1175S}; \citealt{Kenney2015AJ....150...59K}), or gas accretion (\citealt{Bournaud2005AA...438..507B}).   These disturbances may manifest themselves as asymmetries in the gas distribution, velocity field, and global spectra.  More than half of isolated spirals show some degree of asymmetry in their \HI\ spectra (e.g., \citealt{Richter1994_asy, HI_asymmetry_Haynes, Matthews1998AJ....116.1169M}).  Even after accounting for noise-induced effects, $\sim 40$\% of galaxies show asymmetry features \citep{Watts2020MNRAS.492.3672W}.  Perturbations can also arise internally, for instance through feedback from stellar processes (\citealt{Ashley2017AJ....153..132A}) or powerful active galactic nuclei (\citealt{Morganti2017FrASS...4...42M}).

In this work, we develop a new method to measure the central velocity, total flux, line width, asymmetry, and intensity concentration of a global \HI\ profile. The method is robust and efficient, and it can be easily automated for large surveys.  We calibrate our newly defined line width against rotation velocity derived from rotation curves. We use the line widths, in conjunction with radii estimated from the empirical relation between \HI\ size and \HI\ mass, to estimate galaxy dynamical masses.  Our sample is defined in Section~\ref{sec:sample}. Section~\ref{sec:method} describes our methodology, including corrections to and uncertainties of the measurements. Section~\ref{sec:lines} calibrates our line widths and dynamical masses with literature results. Our main conclusions are summarized in Section~\ref{sec:sum}.

\section{Sample Compilation}
\label{sec:sample}

The strategy behind our approach is to directly calibrate, for a set of nearby reference galaxies, the line widths measured from single-dish, integrated \HI\ spectra with accurate rotation velocities derived from well-resolved rotation curves.  Our primary sample derives from the work of \citet{Wang+2016HI-size-mass}, who compiled data from 15 \HI\ interferometric projects for 549 nearby galaxies \citep{BroeilsRhee1997, VerheijenSancisi2001UM_HI,  Swaters2002AA...390..829S, Noordermee2005whisp(sa), Begum2008MNRAS.386.1667B, Walter2008AJ....136.2563W, Chung2009AJ....138.1741C, Kovac2009MNRAS.400..743K, Hunter2012AJ....144..134H, Kreckel2012AJ....144...16K, Serra2012MNRAS.422.1835S, Serra2014MNRAS.444.3388S, Wang2013MNRAS.433..270W, Lelli2014AA...566A..71L, Martinsson2016AA...585A..99M, Koribalski2018MNRAS.478.1611K}, covering over 5 orders of magnitudes in \HI\ mass and more than 10 optical magnitudes. \citet{Wang+2016HI-size-mass} only selected galaxies with sizes at least twice as large as the synthesis beam to ensure that the \HI\ is well-resolved. The \HI\ diameter is measured as the major axis of the isophote at a surface density level of 1~\msun~pc$^{-2}$.  After removing redundant sources, the initial sample contains 469 unique galaxies. We apply two further cuts, as follows.

Making extensive use of the NASA/IPAC Extragalactic Database (NED)\footnote{{\url http://ned.ipac.caltech.edu/forms/Searchspectrum.html}.}, we located published single-dish \HI\ spectra for 387 of the 469 galaxies from the initial sample. To avoid missing flux from single-pointing observations, we require that the beam of the telescope used in the observation be larger than the diameter of the \HI\ disk.  If more than one spectrum is available, we choose the one taken with the smallest beam in order to minimize potential contamination from nearby sources.  When multiple spectra are available from the same telescope, for the sake of homogeneity we give preference to \cite{Haynes+2011AJ....142..170H} for data from the Arecibo 305~m and \cite{Koribalski2004} for data from the Parkes 65~m.  All else being equal, we choose spectra with the highest velocity resolution. Taking these considerations into account reduces the sample to 346 galaxies.  Careful inspection reveals that the spectra of six objects (DDO~69, NGC~2976, UGC~6969, UGC~8508, UGC~9177, and VII~Zw 403; see Appendix~A) suffer from strong radio-frequency interference or other obvious problems, and they were discarded.  The spectra of another six galaxies (AGC~190187, NGC~3972, NGC~4010, NGC~4501, UGC~448, and UGC~6983; see Appendix~A) are also contaminated by artifacts, but we were able to salvage them by masking out the contaminated channels.  This leaves us with a final sample of 340 galaxies.

Next, we searched the literature for the availability of robust rotation velocities ($V_{\rm rot}$) derived from spatially resolved observations. The observations employ diverse techniques and kinematic tracers.  The majority of the measurements come from \HI\ position-velocity diagrams, rotation curves, or velocity fields (\citealt{Guhathakurta1988AJ.....96..851G}; \citealt{Verheijen2001ApJ...563..694V}; \citealt{Rhee2005JASS...22...89R}; \citealt{Noordermeer2007MNRAS.376.1513N}; \citealt{Begum2008MNRAS.386..138B}; \citealt{Trachternach2008AJ....136.2720T}; \citealt{Swaters2009AA...493..871S}; \citealt{Lelli2014AA...566A..71L}; \citealt{denHeijer2015AN....336..284D}; \citealt{Oh2015AJ....149..180O}; \citealt{Koribalski2018MNRAS.478.1611K}), with preference given to the set of high-quality \HI\ data from \citet{Lelli2016AJ....152..157L} that probe the flat part of the rotation curve.  Measurements of $V_{\rm rot}$ based on spatially resolved 21~cm observations usually are considered reliable when the \HI\ distribution is more extended than 4 times the major axis of the synthesis beam of the interferometer.  Only optical emission-line rotation curves could be used for some galaxies \citep{Rubin1999AJ....118..236R}, while the rotation velocities of others were obtained from a combination of optical rotation curves and \HI\ velocity maps \citep{Martinsson2013AA...557A.130M}.  A minority of the sample are early-type galaxies for which only stellar kinematics are available; their rotation velocities were derived through axisymmetric dynamical models \citep{Cappellar2013MNRAS.432.1709C}.  In total, 269 galaxies have reliable published values of $V_{\rm rot}$, and these constitute our final sample, which is summarized in Table~\ref{tab:basic}.  The uncertainties of $V_{\rm rot}$, if available, are quoted directly from the original literature sources; otherwise, we assume a typical uncertainty of 5 \kms, which is the median value of the galaxies with published uncertainties.

The data collected for our analysis derive from a large number of heterogeneous sources, and their flux intensity units need to be homogenized.  The flux densities for 80 galaxies were published in units of antenna temperature.  For the data from \citet{1978ApJ...223..391D} and \citet{TifftCocke1988ApJS...67....1T}, we follow \citet{TifftCocke1988ApJS...67....1T}: $F= 0.0042T_{\mathrm{0.01}}/(0.4665+0.003595\delta-4.98\times 10^{\mathrm{-5}} \delta^2)\,{\rm Jy},$ where $\delta$ is declination of the object in units of radians and $T_{\mathrm{0.01}} = 0.01\, {\rm K}$ is the antenna temperature.  In the case of data from \citet{Bieging1978AA....64...23B} and \citet{1979ApJ...227..776K}, $F= T/G$, where $T$ is the antenna temperature in units of K and the gain $G = 8.5\,{\rm K~Jy}^{-1}$ for the Arecibo 305~m telescope \citep{1979ApJ...227..776K} and $G = 0.74\,{\rm K~Jy}^{-1}$ for the Effelsberg 100~m telescope \citep{Bieging1978AA....64...23B}.

The spectral resolution of the data---a combination of instrumental resolution and smoothing, if applied---covers a wide range from 2.8\,\kms\ to 30.0\,\kms, with a median value of 11.0\,\kms.  A significant fraction of the spectra (145/269) retrieved from NED were scanned and then resampled in velocity space.  We verified that the scans used velocity steps smaller than the channel width of the original spectra, and thus the scanning process did not significantly affect the original velocity resolution. Hence, we only consider the channel width and smoothing effects in the original spectrum when correcting for the instrumental resolution.

\begin{deluxetable*}{crrccrccD@{$\pm$}DrD@{$\pm$}Drrrcl} 
\centering
\small\addtolength{\tabcolsep}{-3pt}
\tabletypesize{\footnotesize}
\tablecolumns{14}
\tablewidth{0pt} 
\tablecolumns{15}
\tablecaption{Basic Properties of the Sample}
\tablehead{
\colhead{Galaxy} &
\colhead{R. A.} &
\colhead{Decl.} &
\colhead{$T$} &
\colhead{$z$} &
\colhead{$D_L$} &
\colhead{$V_{\rm rot}$} &
\colhead{$i_{\rm ref}$} &
\multicolumn4c{$q$} &
\colhead{\mstar} &
\multicolumn4c{$i$} &
\colhead{$V_{\rm c, ref}$} &
\colhead{$F_{\rm ref}$} &
\colhead{$v_{\rm inst}$} &
\colhead{log \mtwo} &
\colhead{Ref.}
\\ 
\colhead{} &      
\colhead{($^{\circ}$)} &
\colhead{($^{\circ}$)} &
\colhead{Type} &  
\colhead{} &  
\colhead{(Mpc)} &    
\colhead{(\kms)} &     
\colhead{($^{\circ}$)} &
\multicolumn4c{} &
\colhead{(\msun)} &
\multicolumn4c{($^{\circ}$)} &
\colhead{(\kms)} &  
\colhead{(Jy \kms)} & 
\colhead{(\kms)} &   
\colhead{(\msun)} & 
\colhead{}
\\
\colhead{(1)} &
\colhead{(2)} &
\colhead{(3)} &
\colhead{(4)} &
\colhead{(5)} &
\colhead{(6)} &
\colhead{(7)} &
\colhead{(8)} &
\multicolumn4c{(9)} &
\colhead{(10)}&
\multicolumn4c{(11)} &
\colhead{(12)}&
\colhead{(13)}&
\colhead{(14)}&
\colhead{(15)}&
\colhead{(16)}
}
\decimals
\startdata
UGC~1281 & 27.38212 & 32.58783 & 7 & 0.000520 & 5.5 & 55$\pm$3 & 90 & 0.22 & 0.03 & 7.6 & 82 & 7 & 156.9 & 38.1 & 11.0 & 10.5$\pm$0.6 & 1,19,47 \\
UGC~2023 & 38.32550 & 33.49042 & 9 & 0.001965 & 10.1 & 59$\pm$5 & \nodata & 0.97 & 0.11 & 8.6 & 13 & 27 & 603.5 & 16.7 & 11.0 & 10.9$\pm$0.6 & 2,19,46 \\
UGC~2034 & 38.42883 & 40.52867 & 9 & 0.001928 & 10.1 & 47$\pm$5 & \nodata & 0.84 & 0.10 & 9.5 & 33 & 11 & 578.2 & 31.4 & 11.0 & \nodata & 2,19,\nodata \\
UGC~2053 & 38.62212 & 29.74983 & 9 & 0.003432 & 11.8 & 99$\pm$5 & \nodata & 0.55 & 0.08 & 7.0 & 60 & 8 & 1025.5 & 15.6 & 11.0 & \nodata & 2,19,\nodata \\
UGC~3371 & 89.14992 & 75.31719 & 9 & 0.002722 & 12.8 & 86$\pm$5 & \nodata & 0.76 & 0.09 & 8.8 & 41 & 9 & 814.4 & 31.6 & 11.0 & 11.0$\pm$0.6 & 2,19,54 \\
UGC~3817 & 110.68533 & 45.10853 & 9 & 0.001458 & 8.7 & 45$\pm$5 & \nodata & 0.51 & 0.07 & 8.9 & 60 & 6 & 438.2 & 10.2 & 11.0 & \nodata & 2,19,\nodata \\
UGC~4173 & 121.79187 & 80.12681 & 9 & 0.002869 & 16.8 & 57$\pm$5 & \nodata & 0.47 & 0.04 & 6.3 & 67 & 6 & 860.5 & 29.5 & 11.0 & \nodata & 2,19,\nodata \\
UGC~4278 & 123.49545 & 45.74215 & 6 & 0.001801 & 10.5 & 91$\pm$4 & 90 & 0.14 & 0.01 & 8.1 & 90 & 5 & 557.0 & 46.6 & 13.0 & 10.9$\pm$0.6 & 1,33,46 \\
UGC~4325 & 124.83554 & 50.00961 & 9 & 0.001748 & 10.1 & 90$\pm$2 & 41 & 0.54 & 0.03 & 9.5 & 58 & 5 & 517.0 & 21.9 & 6.6 & 11.3$\pm$0.1 & 1,22,48 \\
UGC~4499 & 129.42287 & 51.65238 & 8 & 0.002305 & 13.0 & 72$\pm$2 & 50 & 0.44 & 0.04 & 8.6 & 65 & 5 & 691.0 & 25.6 & 11.0 & 10.7$\pm$0.6 & 1,19,46 \\
\enddata
\tablecomments{Col. (1): Galaxy name. Cols. (2)--(3): Equatorial coordinates (J2000). Col. (4): Morphological type index from HyperLeda. Col. (5): Optical redshift from NED. Col. (6): Luminosity distance from \citet{Wang+2016HI-size-mass}. Col. (7): Deprojected rotation velocity from the literature.  Col. (8): Inclination angle from the literature. Col. (9) Optical axis ratio. Col. (10): Stellar mass derived in this work.  Col. (11): Inclination angle derived in this work. Col. (12): Central velocity. Col. (13): Total integrated \HI\ flux. Col. (14): Instrumental velocity resolution. Col. (15): Total mass inside $r_{200}$, derived from mass model. Col. (16): References for Cols. (7)--(8), Cols. (12)--(14), and Col. (15).  (Table~1 is published in its entirety in machine-readable format. A portion is shown here for guidance regarding its form and content.)}
\tablerefs{
 (1) \citealt{Lelli2016AJ....152..157L};
 (2) \citealt{Swaters2009AA...493..871S};
 (3) \citealt{Koribalski2018MNRAS.478.1611K};
 (4) \citealt{Lelli2014AA...566A..71L};
 (5) \citealt{denHeijer2015AN....336..284D};
 (6) \citealt{Martinsson2013AA...557A.130M};
 (7) \citealt{Trachternach2008AJ....136.2720T};
 (8) \citealt{Leroy2008AJ....136.2782L};
 (9) \citealt{VerheijenSancisi2001UM_HI};
(10) \citealt{Verheijen2001ApJ...563..694V};
(11) \citealt{Rubin1999AJ....118..236R};
(12) \citealt{Guhathakurta1988AJ.....96..851G};
(13) \citealt{Noordermeer2007MNRAS.376.1513N};
(14) \citealt{Cappellar2013MNRAS.432.1709C};
(15) \citealt{Oh2015AJ....149..180O};
(16) \citealt{Rhee2005JASS...22...89R};
(17) \citealt{Begum2008MNRAS.386..138B};
(18) \citealt{Springob+2005};
(19) \citealt{TifftCocke1988ApJS...67....1T};
(20) \citealt{Huchtmeier1986AAS...63..323H};
(21) \citealt{Koribalski2004}; 
(22) \citealt{1978ApJ...223..391D};
(23) \citealt{Haynes+2011AJ....142..170H};
(24) \citealt{1979ApJS...40..527P};
(25) \citealt{2014MNRAS.443.1044M};
(26) \citealt{Richter1991AAS...87..425R};
(27) \citealt{1975ApJ...198..527S};
(28) \citealt{Shostak1978AA....68..321S};
(29) \citealt{Courtois+2009AJ....138.1938C};
(30) \citealt{1987MNRAS.224..953S};
(31) \citealt{Rots1980AAS...41..189R};
(32) \citealt{1983AJ.....88..272H};
(33) \citealt{Huchtmeier2005AA...435..459H};
(34) \citealt{1987ApJS...63..515L};
(35) \citealt{1979ApJ...227..776K};
(36) \citealt{Haynes1984AJ.....89..758H};
(37) \citealt{1986AJ.....91..705B};
(38) \citealt{1995ApJS...96....1M};
(39) \citealt{1978ApJ...226..770P};
(40) \citealt{Bieging1978AA....64...23B};
(41) \citealt{1986AJ.....91..732B};
(42) \citealt{1995AJ....110..581M};
(43) \citealt{Matthews1998AJ....116.1169M};
(44) \citealt{Bottinelli1993AAS..102...57B};
(45) \citealt{Roberts1978AJ.....83.1026R};
(46) \citealt{Li2019MNRAS.482.5106L};
(47) \citealt{Forbes2018MNRAS.481.5592F};
(48) \citealt{Adams2014ApJ...789...63A};
(49) \citealt{deBlok2008AJ....136.2648D};
(50) \citealt{Chemin2011AJ....142..109C};
(51) \citealt{Oh2011AJ....142...24O};
(52) \citealt{Korsaga2019MNRAS.482..154K};
(53) \citealt{Truong2017PhDT........28T};
(54) \citealt{vandenBosch2001MNRAS.325.1017V}.
}
\label{tab:basic}
\end{deluxetable*}

\section{Methodology}
\label{sec:method}

\subsection{Deriving the Curve-of-Growth}
\label{subsec:gc}

We develop a new method to measure the properties of an emission line by constructing its curve-of-growth (CoG).  The CoG is calculated by integrating the flux intensity as a function of velocity from the center of the line outward to both sides of the profile. If the baseline has been properly subtracted and the noise distribution is approximately symmetric around zero, the CoG should rise continuously until there is no net signal to be gained.  The flux at which the CoG converges defines the total flux, and the profile of the CoG delineates the variation of velocity width as a function of fractional line flux.  Moreover, the CoG of each side of the line yields information on the degree of line asymmetry.  The CoG method allows the possibility of full automation, an important consideration for application to large \HI\ surveys.

We first mask out strong contaminating features, such as obvious radio frequency interference or foreground emission or absorption from the Milky Way, which usually can be recognized by their very high flux densities (e.g., $\gtrsim 1$ Jy).  A total of 78 spectra have nonzero or obviously poorly subtracted baselines, and for these we refit and subtract their baselines. We estimate the noise level of the spectrum by examining the distribution of flux densities of all the channels.  With the baseline properly subtracted, the main peak of the flux intensity distribution is centered at zero.  While the positive side of the distribution may contain real signal from the galaxy, the negative side of the distribution should reflect the noise in the spectrum.  Assuming the noise distribution to be symmetric about zero, we mirror the negative side of flux intensity distribution to the positive side and fit the resulting distribution with a Gaussian function.  The standard deviation of the best-fit Gaussian is then taken to be the $1\,\sigma$ rms of the spectrum.  This method mitigates the influence of galaxy signal on the rms measurement, and it has been widely used in the literature (e.g., \citealt{LiHo2011ApJS..197...22L}; \citealt{Serra2012PASA...29..296S}).

We search for emission signal within a range $\pm 500$ \kms\, around the optical central velocity. We first select segments of at least three consecutive channels whose flux intensities are above 0.  We calculate the mean flux intensity of each segment and the maximum value of the mean flux intensity for all segments. Only segments whose mean flux intensity exceeds 0.7 times\footnote{This is a purely empirical criterion determined after tests with different thresholds.} the maximum mean flux intensities are considered as possible emission signal.  This ensures that the selected channel segments span a reasonable range in velocity and flux intensity. 

\begin{figure*}[t]
\epsscale{1.1}
\plotone{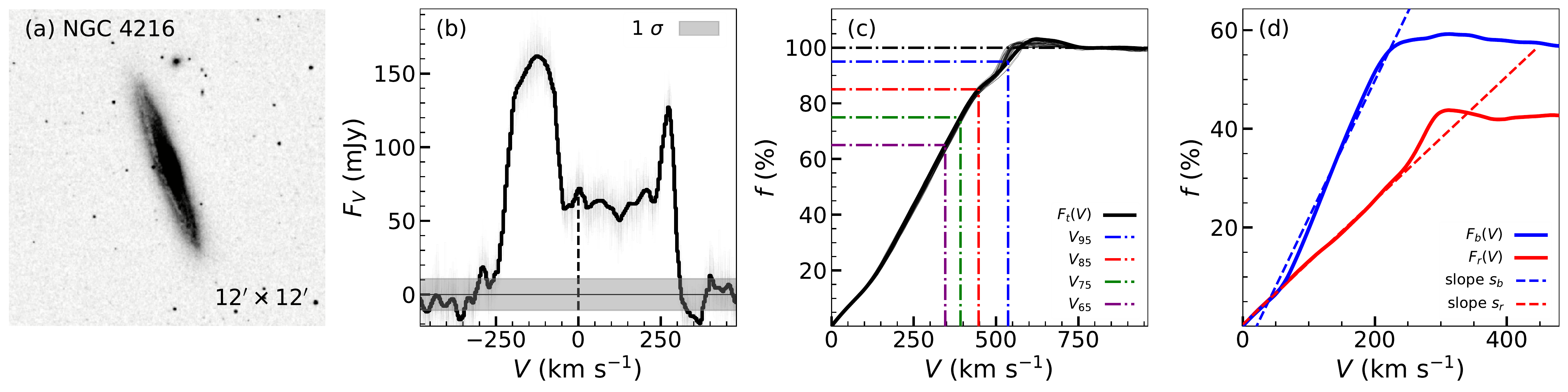}
\plotone{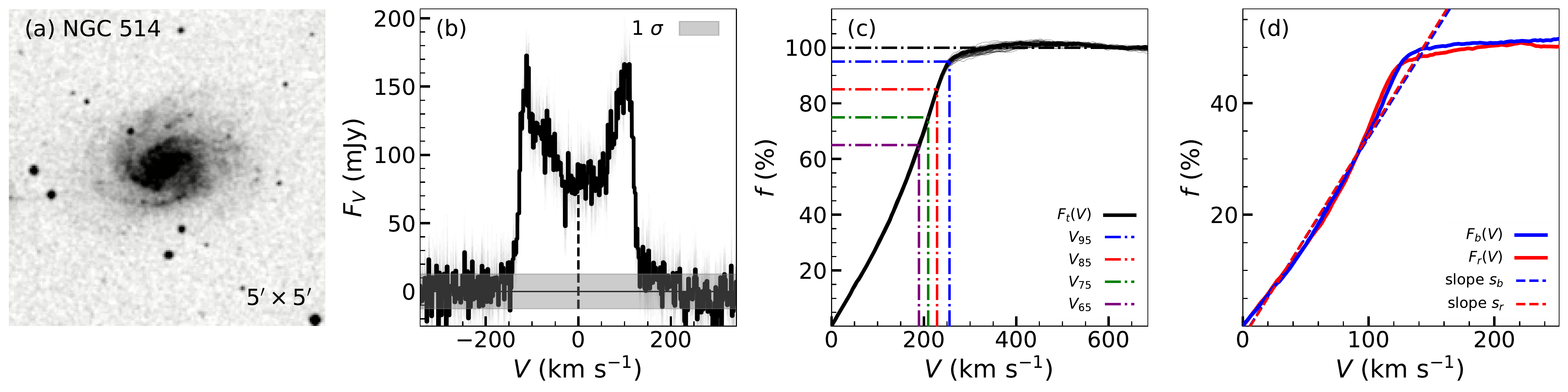}
\caption{Illustration of our analysis technique for (top) NGC~4216, a galaxy with an asymmetric \HI\ profile, and (bottom) NGC~514, a galaxy with a symmetric \HI\ profile.  Panel (a) shows the optical image, whose size is given on the lower-right corner.  Panel (b) displays the \HI\ spectrum (thick black line) on a velocity scale centered in the rest frame; the grey shaded region shows the $1\,\sigma$ uncertainty of the flux intensity.  Panel (c) presents the normalized CoG for the total spectrum $f = F_t(V)/F$ (black solid).  The dot-dashed lines mark the line widths at 65\% ($V_{65}$; purple), 75\% ($V_{75}$; green), 85\% ($V_{85}$; red), and 95\% ($V_{95}$; blue) of the integrated line flux, which converges at $f=100\%$ (black).  Panel (d) shows the normalized CoG for the blue side of the spectrum $f = F_b(V)/F$ (solid blue) and the red side of the spectrum $f = F_r(V)/F$ (solid red). The linear fits of the rising part of the CoG are shown as blue (slope $s_b$) and red (slope $s_r$) dashed lines.  The thin black lines in panels (b) and (c) are the results from 50 sets of Monte Carlo simulations.}
\label{fig:mExample}
\end{figure*}

\begin{figure*}
\centering
\epsscale{1.0}
\subfigure{\includegraphics[width=0.9\textwidth]{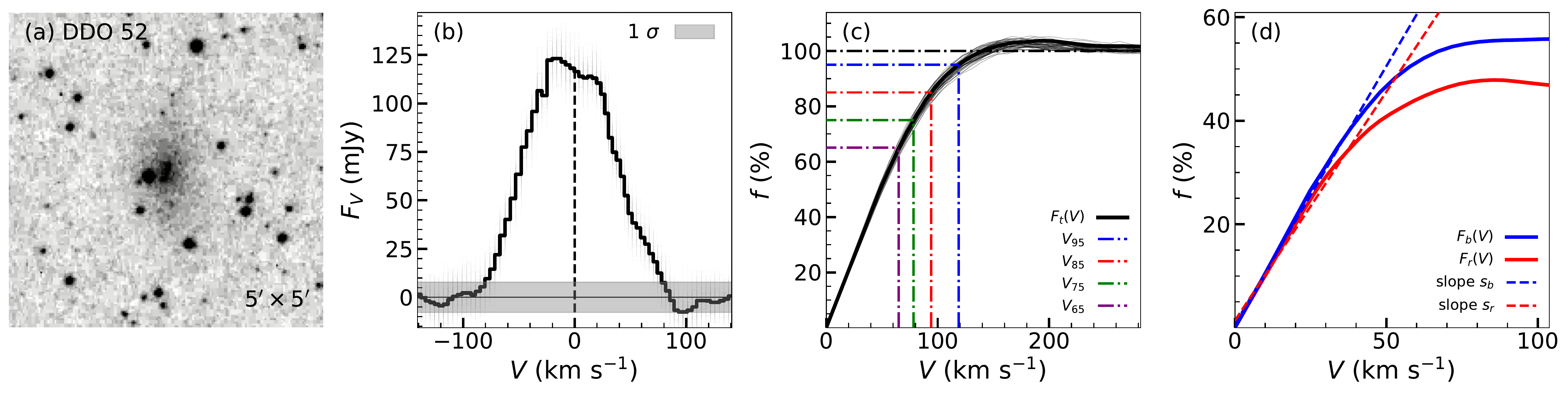}}
\subfigure{\includegraphics[width=0.9\textwidth]{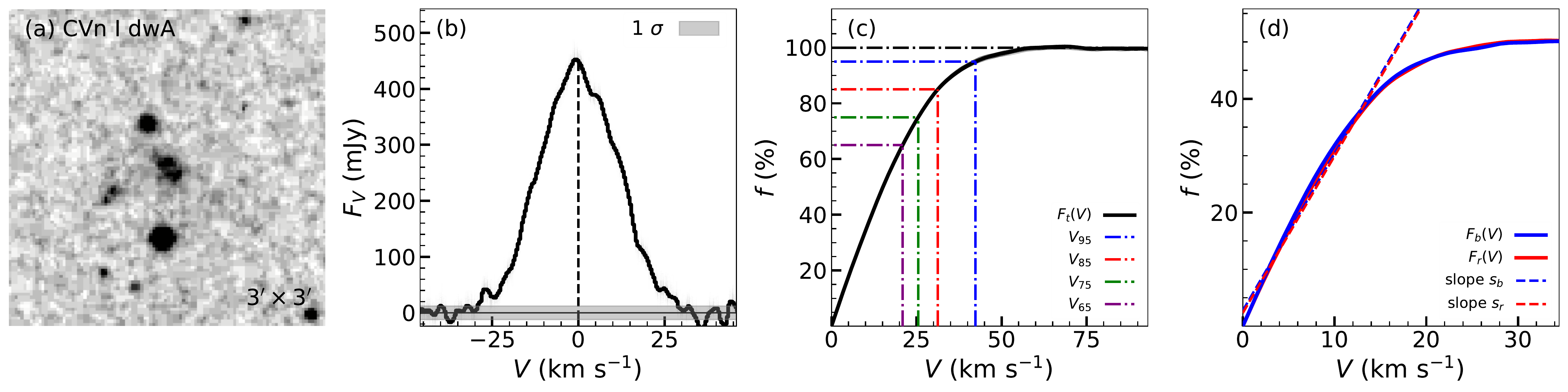}}
\subfigure{\includegraphics[width=0.9\textwidth]{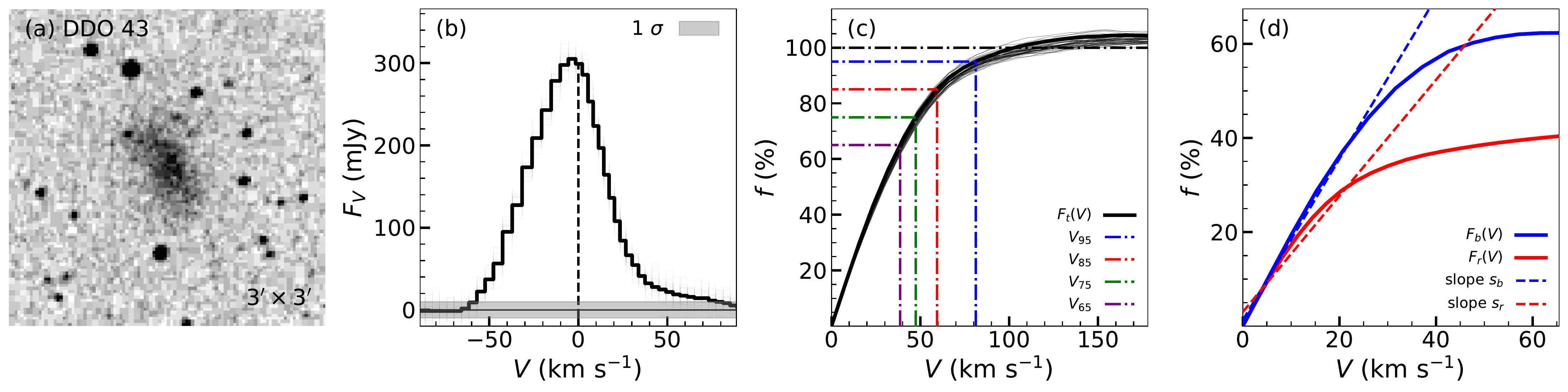}}
\subfigure{\includegraphics[width=0.9\textwidth]{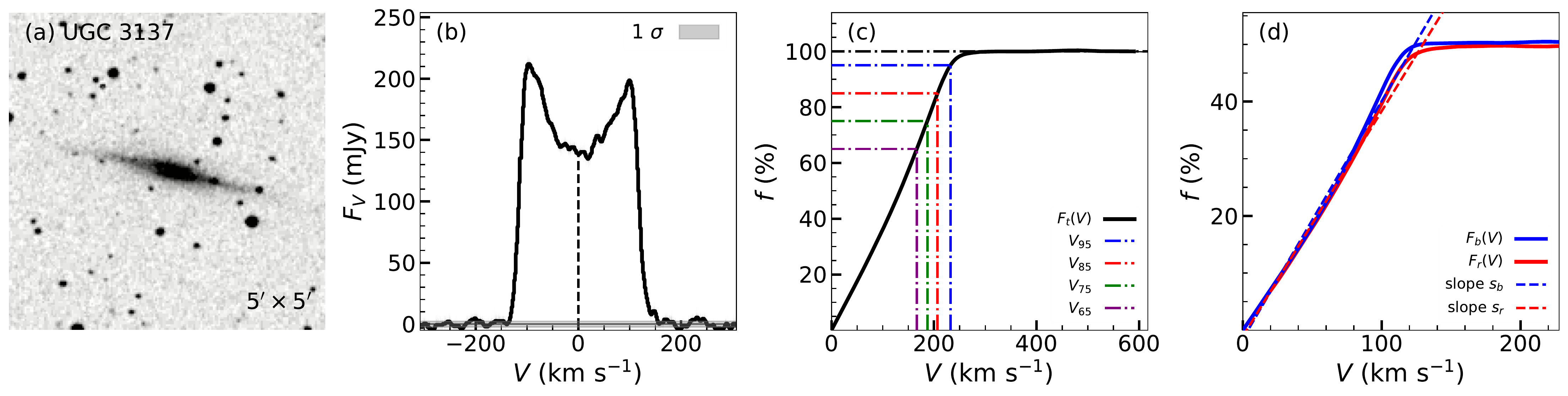}}
\subfigure{\includegraphics[width=0.9\textwidth]{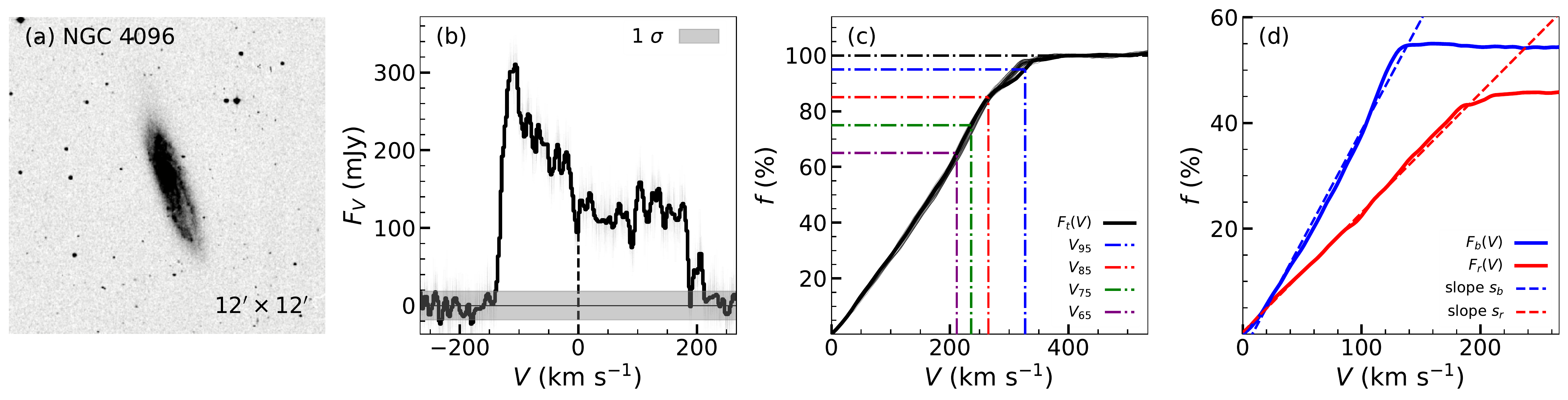}}
\caption{Similar to Figure~\ref{fig:mExample}. These five examples illustrate the range of typical \HI\ profiles: flat-Gaussian (DDO~52), symmetric Gaussian (CVn~I~dwA), asymmetric Gaussian (DDO~43), symmetric double-horn (UGC~3137), and asymmetric double-horn (NGC~4096).  The spectra and CoGs of the full sample are available in the electronic version of the paper.}
\label{appfig:CoGexample}
\end{figure*}

The lowest and highest velocities of the segments enclose $N$ channels, which we use to derive a flux intensity-weighted central velocity $V_c$.  Using a total velocity range that spans $3N$ channels, the CoG for the blue side, the red side, and the whole region (both sides combined) are calculated as

\begin{equation}
\begin{split}
F_b(V) &=\int^{V_c}_{V_c-V} F_V dV,\\
F_r(V) &=\int_{V_c}^{V_c+V} F_V dV,\\
F_t(V) &= F_b(V) + F_r(V),
\end{split}
\label{equ:Vt}
\end{equation}

\noindent
where $F_V$ is the flux intensity at velocity $V$.  Figure~\ref{fig:mExample} illustrates the normalized CoG for two example galaxy spectra, one asymmetric (NGC~4216) and the other symmetric (NGC~514).  Figure~\ref{appfig:CoGexample} shows the CoGs for five representative \HI\ spectra that span the range of typical line profiles: flat-Gaussian, symmetric Gaussian, asymmetric Gaussian, symmetric double-horn, and asymmetric double-horn.  The spectra and CoGs for the full sample are available in the electronic version of the paper.

\subsection{Measuring Line Parameters}
\label{subsec:PresStV}

The CoG yields several useful physical quantities in a straightforward and well-defined manner.  We focus on four sets of parameters that measure (1) total flux, (2) velocity widths, (3) line asymmetry, and (4) concentration.

\begin{figure}
\epsscale{0.9}
\plotone{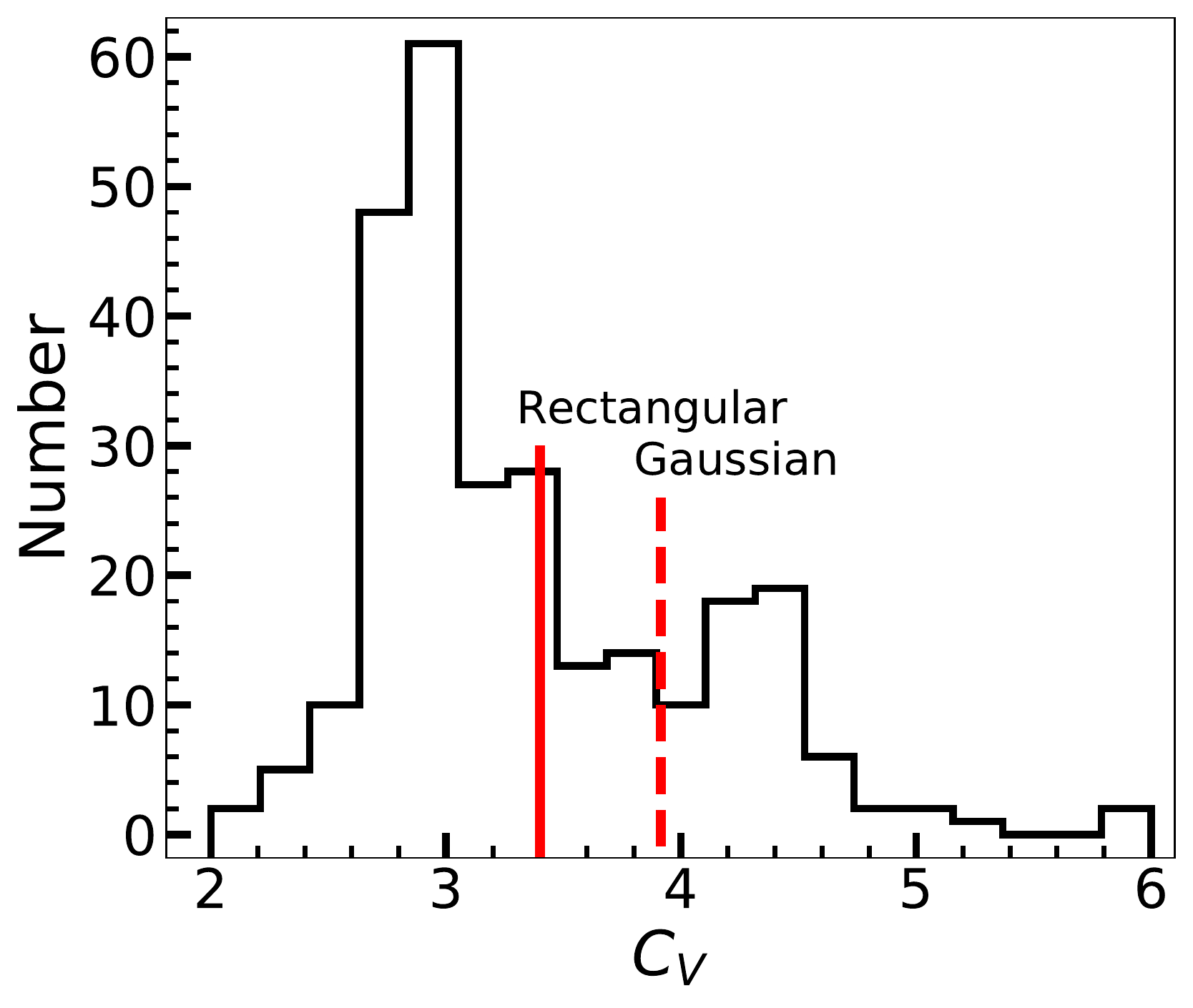}
\caption{Distribution of \HI\ concentration, $C_V = V_{85}/V_{25}$. The red solid and dashed lines mark the typical concentration of a rectangular ($C_V = 3.4$) and Gaussian ( $C_V = 3.9$) profile, respectively.}
\label{fig:cctn}
\end{figure}

\begin{itemize}
\item{The total flux of the line, $F$, is taken as the median integrated flux on the flat part of the CoG.  The median integrated flux is more stable than the mean or the last integrated flux of the CoG.}

\item{In lieu of the traditional measures of line width, such as $W_{20}$ or $W_{50}$ (e.g., \citealt{Haynes1984AJ.....89..758H}; \citealt{Springob+2005}; \citealt{Courtois+2009AJ....138.1938C}), we use the CoG, $F_t(V)$, to define velocity widths that enclose different percentages of the total line flux.  Our technique is illustrated graphically in Figures~\ref{fig:mExample} and \ref{appfig:CoGexample}.  For example, $V_{85}$, the velocity width that captures 85\% of the total flux, is simply the velocity distance along the $x$-axis of the CoG that corresponds to $f = F_t(V)/F = 0.85$.  In addition to $V_{85}$, we will later investigate $V_{65}$, $V_{75}$, and $V_{95}$, which are defined in a similar manner.}

\item{The degree of symmetry or lack thereof of the line profile carries useful information about the spatial distribution and kinematics of the gas.  Following the definition of \citet{HI_asymmetry_Haynes}, we define the flux asymmetry parameter as ratio of the fluxes on the two sides of the profile around the central velocity: 

\begin{equation}
\begin{split}
A_F & = F_b/F_r, \ \mathrm{if}\  F_b \ge F_r \\
    & = F_r/F_b, \ \mathrm{if}\  F_b  <  F_r,\\
\end{split}
\label{equ:af}
\end{equation}

\noindent
where $F_{b}$ and $F_{r}$ are integrated fluxes of the blue and red sides of the profile, respectively.  We introduce a second, new asymmetry parameter, one that is sensitive to the flux distribution on each side of the profile, based on the slope of the CoG on the red and blue sides of the profile.  For each side of the profile, we fit a straight line to the rising part of the CoG.  From the slope of the fit on the blue ($s_{b}$) and red side ($s_{r}$), we define

\begin{equation}
\begin{split}
A_C   & = s_{b}/s_{r}, \ \mathrm{if}\  s_{b} \ge s_{r} \\
      & = s_{r}/s_{b}, \ \mathrm{if}\  s_{b}  <  s_{r}.\\
\end{split}
\label{equ:ac}
\end{equation}

\noindent
For the examples given in Figure~\ref{fig:mExample}, the obviously asymmetric \HI\ profile of NGC~4216 has $A_F = 1.36$ and $A_C=2.21$, while the more symmetric profile of NGC~514 yields $A_F = 1.06$ and $A_C=1.02$.}

\item{In image analysis, the concentration of the light distribution of an extended source is commonly characterized by the ratio of two metric radii that enclose different fractions of the total flux (e.g., \citealt{Strateva2001AJ....122.1861S}; \citealt{Wang2013MNRAS.433..270W}; \citealt{YuSY2019ApJ...871..194Y}).  In a similar spirit, to characterize the degree of concentration of the line profile, we define the parameter 

\begin{equation}
C_V   = V_{85}/V_{25}, 
\label{equ:cv}
\end{equation}

\noindent
where $V_{25}$ and $V_{85}$ are the line widths enclosing 25\% and 85\% of the total flux of the CoG, respectively.  A Gaussian profile has $C_V = 3.9$, while for a boxcar profile $C_V = 3.4$. Depending on the depth of the central trough, a double-horn profile may have lower values of $C_V$ than a boxcar or Gaussian profile. The distribution of $C_V$ for our sample is shown in Figure~\ref{fig:cctn}: 64\% (171/269) of the profiles have $C_V \leq 3.4$, indicative of their double-horn profiles.}
\end{itemize}

\subsection{Estimates of Signal-to-noise Ratio and Uncertainties}
\label{subsec:error}

The signal-to-noise ratio of the spectrum is calculated as

\begin{equation}
{\rm SNR} = \frac{1000F}{\sigma \, \delta V \sqrt{N_{\mathrm{chan}}}\sqrt{N_{\mathrm{smo}}}},
\label{equ:snr}
\end{equation}

\noindent
where $F$ is the total flux in units of Jy~km~s$^{-1}$, $\sigma$ is the rms of the spectrum in units of mJy, $\delta V$ is the channel width in units of km~s$^{-1}$, $N_{\rm chan}=3N$ is the number of channels included in the CoG calculation, and $N_{\rm smo}$ is the window length used to smooth the spectrum. If no smoothing is applied, $N_{\rm smo}=1$.

We consider two sources of uncertainty, both statistical and systematic, for our measured parameters.  The typical uncertainties are summarized in Table~\ref{tab:uncert}.  To estimate statistical uncertainty, we perform Monte Carlo simulations by generating for each spectrum a set of 50 mock spectra by adding to the original spectrum random noise following a Gaussian distribution with standard deviation equivalent to the rms.  We analyze the mock spectra in the same way as the real spectra, and we adopt the standard deviation of the distribution of each parameter as its statistical uncertainty. Because the rms of the mock spectra is $\sim \sqrt{2}$ times larger than the rms of the original spectra, these uncertainty estimates are conservative.  

\begin{figure*}
\epsscale{1.2}
\plotone{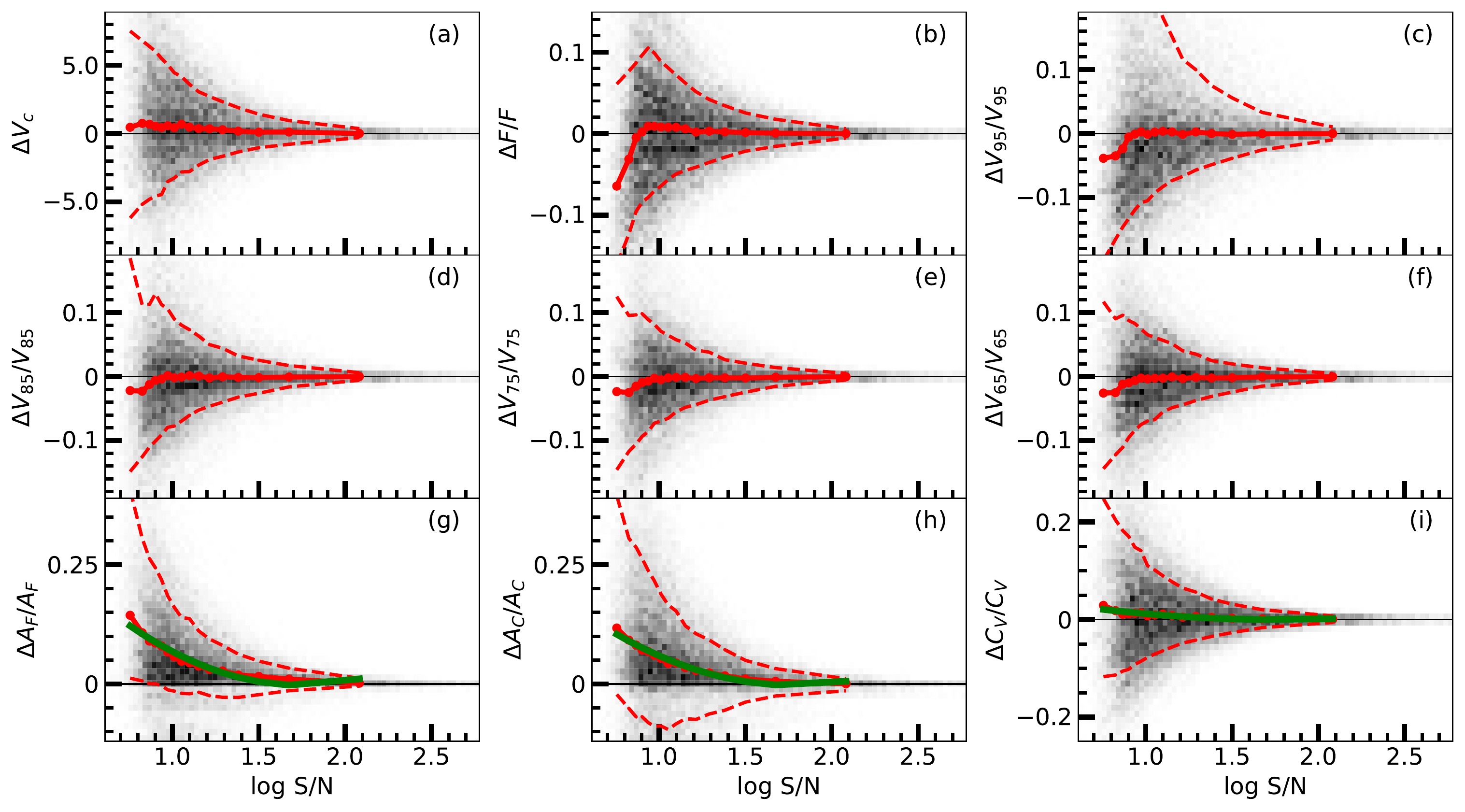}
\caption{Simulations using mock spectra to test the sensitivity of our parameter measurements to variations in SNR.  The mock spectra were generated with five types of representative profile shapes (flat-Gaussian, symmetric Gaussian, asymmetric Gaussian, symmetric double-horn, and asymmetric double-horn). The panels plot (a) absolute deviations for central velocity $V_c$, and fractional deviations for (b) flux $F$, (c)--(f) velocity widths $V_{95}$, $V_{85}$, $V_{75}$, and $V_{65}$, (g) asymmetry parameter $A_F$, (h) asymmetry parameter $A_C$, and (i) concentration parameter $C_V$.  The black horizontal line denotes zero offset. The red solid lines show the median, while the dashed lines show the 16th and 84th percentiles of the distribution. Green curves in panels (g), (h), and (i) are polynomial fits to the median distribution of asymmetry and concentration parameters.}
\label{fig:test}
\end{figure*}

Several sources of systematic uncertainty need to be considered. To evaluate the effect of noise on our parameter measurements, we perform a series of experiments to quantify the performance of our CoG method at different levels of SNR.  We select five spectra with shapes that are representative of the typical range of profiles observed in galaxies: flat-Gaussian (UGC~1281), symmetric Gaussian (UGC~2053), asymmetric Gaussian (NGC~1156), symmetric double-horn (UGC~11861), and asymmetric double-horn (UGC~4278). We fit these observed profiles with a busy function \citep{WestmeierBF2014MNRAS.438.1176W} to generate five model (noise-free) spectra.  Each model is sampled with a velocity resolution of 5.5 \kms, the channel spacing of the ALFALFA survey, and we generate $10^4$ mock spectra for each model by adding random noise to simulate a range of SNRs from 5 to 600. The mock data of each model spectrum share a similar SNR distribution. Our measurement technique, described in Sections~\ref{subsec:gc} and \ref{subsec:PresStV}, is then applied to the mock spectra to measure the nine primary parameters 
($V_c, F, V_{65}, V_{75}, V_{85}, V_{95}, A_F, A_C$, and $C_V$).  We then compare the recovered parameters with the known input values, as a function of SNR.  For ${\rm SNR} \gtrsim 10$,  which is the case for more than 90\% of the galaxies in our sample, $V_c$ can be recovered to within $\sim 3$ \kms\ and $F, V_{65}, V_{75}, V_{85}$, and $V_{95}$ to within $\lesssim 10\%$ (Figure~\ref{fig:test}; Table~\ref{tab:uncert}).  These parameters are underestimated systematically and the scatter increases when ${\rm SNR} < 10$, but, with the exception of $V_{95}$, the fractional uncertainties never exceed $\sim 20\%$.  

The asymmetry and concentration parameters behave in a more complicated manner.  Their fractional uncertainties at ${\rm SNR} \gtrsim 10$ are $\lesssim 10\%$, but in the low-SNR regime they increase with decreasing SNR, to $\sim 20\%$ for $A_F$ and $\sim 50\%$ for $A_C$.  Moreover, $A_F$ and $A_C$ become systematically overestimated with decreasing SNR; $C_V$ exhibits a similar, but much less sigificant trend.  To account for these effects, we fit a second-order polynomial to quantify the median offset as a function of SNR (green curves in Figure~\ref{fig:test}),

\begin{equation}
\begin{split}
\Delta A_F/A_F &= 0.12 (\log{\rm SNR})^2-0.44 \log{\rm SNR}+0.38\\
\Delta A_C/A_C &= 0.10 (\log{\rm SNR})^2-0.36 \log{\rm SNR}+0.32\\
\Delta C_V/C_V &= 0.02 (\log{\rm SNR})^2-0.07 \log{\rm SNR}+0.06,\\
\end{split}
\label{equ:asyFit}
\end{equation}

\noindent
and apply the corrections to our measurements.  Unless otherwise noted, we correct the asymmetry and concentration parameters for these systematic deviations.
 
We also estimate the systematic uncertainties caused by our method itself, which principally stems from the range of channels chosen to calculate the CoG. A key parameter is the threshold for the mean flux intensity used to select the channel segments.  We evaluate its impact by exploring a range of choices centered on the default value of 0.7~times the maximum mean flux intensity.  At high SNR, the uncertainties introduced by this factor is totally negligible, but at low SNR, they are comparable to the uncertainties arising from SNR.  

The final uncertainty of each parameter is the quadrature sum of the statistical uncertainties and systematic uncertainties. As in \cite{Springob+2005}, we consider an additional uncertainty of 15\% only for the total flux uncertainty to account for the calibration effects (beam attenuation, pointing, flux calibration, and \HI\ self-absorption).

\begin{deluxetable}{ccccccc}
\tabletypesize{\footnotesize}
\tablecolumns{7}
\tablecaption{Measurement Uncertainties}
\tablehead{
\colhead{Parameter} &
\multicolumn{2}{c}{Systematic (low SNR)} &
\colhead{} &
\multicolumn{2}{c}{Systematic (high SNR)} &
\colhead{Statistical} 
\\
\cline{2-3} \cline{5-6} \\
\colhead{} &
\colhead{Method} &
\colhead{Noise} &
\colhead{} &
\colhead{Method} &
\colhead{Noise}&
\colhead{}  \\ 
\colhead{(1)} &
\colhead{(2)} &
\colhead{(3)} &
\colhead{   } &
\colhead{(4)} &
\colhead{(5)} &
\colhead{(6)} 
}
\startdata
$V_c$     & 2.4  & 12.2 &&  0.1 &"2.5 & 2.2\\
$F$   	  & "5\% & 11\% &&  0\% &"5\% & 1\%\\
$V_{95}$  & 39\% & 47\% &&  0\% &17\% & 2\%\\
$V_{85}$  & 22\% & 26\% &&  0\% &"6\% & 2\%\\
$V_{75}$  & 15\% & 18\% &&  0\% &"5\% & 2\%\\
$V_{65}$  & 15\% & 16\% &&  0\% &"5\% & 2\%\\
$A_F$     & 19\% & 19\% &&  0\% &"7\% & 2\%\\
$A_C$     & 68\% & 62\% &&  0\% &10\% & 5\%\\
$C_V$     & 14\% & 62\% &&  0\% &10\% &4\%\\
\enddata
\tablecomments{Col. (1): Measured parameter. The uncertainty for $V_c$ is absolute, while the uncertainty for the other parameters are fractional. Col. (2): Typical systematic uncertainties due to the method if $5< {\rm SNR} < 10$. Col. (3): Typical systematic uncertainties due to SNR if $5< {\rm SNR} < 10$. Col. (4): Typical systematic uncertainties due to the method if ${\rm SNR}>10$. Col. (5): Typical systematic uncertainties due to SNR if ${\rm SNR}>10$. Col. (6): Typical statistical uncertainties.}
\label{tab:uncert}
\end{deluxetable}

\subsection{Comparison of $V_c$ and $F$ with Literature Measurements}
\label{subsec:CompStVc}

\begin{figure}
\epsscale{1.1}
\plotone{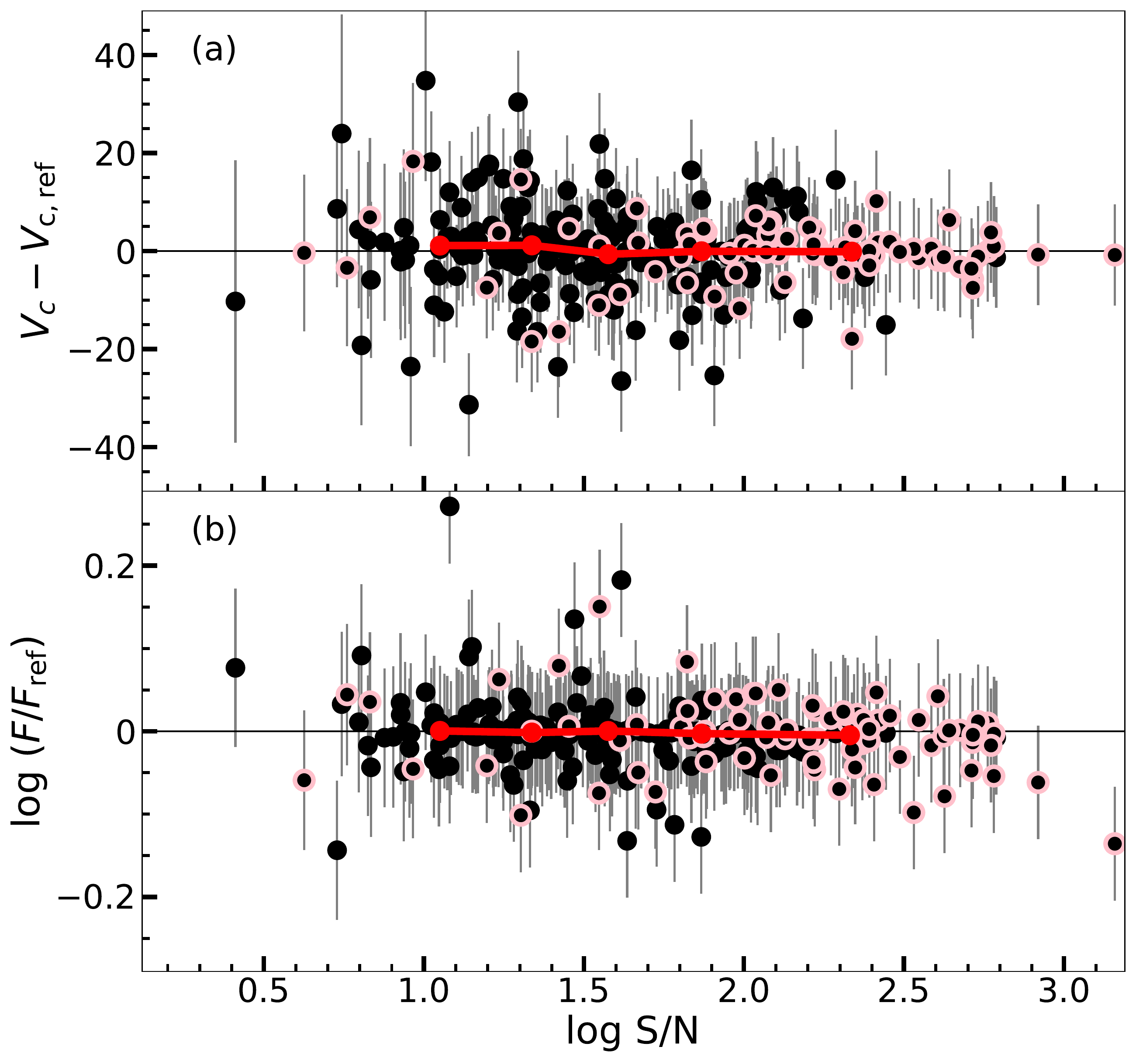}
\caption{Comparison between our measurements of (a) central velocity $V_c$ and (b) total flux $F$ with measurements from the literature as reference, as a function of the SNR of the spectrum. The red solid line shows the median value in each SNR bin. The points highlighted with pink circles were derived from spectra for which we improved the baseline subtraction.}
\label{fig:StVcComp}
\end{figure}

Among the spectral parameters we measure, two of them can be compared in a straightforward manner with literature measurements.  Whereas our values of $V_c$ are flux intensity-weighted, the literature values were typically calculated as the midpoint of the channels whose flux densities drop to 20\% \citep{Richter1991AAS...87..425R} or 50\% \citep{1986AJ.....91..705B} of the mean, or they were evaluated using the peak flux of the line \citep{Matthews1998AJ....116.1169M}. Figure~\ref{fig:StVcComp} shows that both sets of values for $V_c$ are quite consistent for ${\rm SNR} > 100$ ($\Delta V_c = 0.6\pm5.8$ \kms), with the scatter increasing toward ${\rm SNR} < 30$ ($\Delta V_c = -2.4\pm11.9$ \kms).  The integrated flux shows good overall agreement over the full range of SNR ($\Delta \log F =  0.00\pm0.04$ dex).

\section{Linking Line Widths to Physical Quantities}
\label{sec:lines}

\subsection{Rotation Velocities}
\label{subsec:cal}

One of our primary goals is to use the line widths to estimate rotation velocities. Apart from the effect of redshift, which stretches the line width by a factor $(1 + z)$, additional corrections need to be applied for instrumental broadening, turbulence motion, and projection due to inclination (\citealt{Springob+2005}; \citealt{Ho2008ApJS..177..103H}; \citealt{Meyer2008MNRAS.391.1712M}).

\subsubsection{Corrections for Instrumental and Turbulent Broadening}
\label{subsec:LinesCorSimu}

The observed line widths must be corrected for broadening due to instrumental resolution and turbulence motions.  These effects are traditionally corrected with a linear or quadratic subtraction. \citet{Bottinelli1983AA...118....4B} were among the first to suggest a linear correction because the observed profiles of many galaxies can be approximated as a boxcar (reflecting rotation) convolved with a Gaussian function (reflecting additional broadening).  \citet{TullyFouque1985ApJS...58...67T} pointed out that the linear subtraction can be problematic because the boxcar approximation may fail for galaxies with low intrinsic line width. They proposed a linear subtraction for massive galaxies and a quadratic subtraction for low-mass systems, which has became standard practice for broadening correction of the conventionally used line widths $W_{20}$ and $W_{50}$ (e.g., \citealt{VerheijenSancisi2001UM_HI}; \citealt{Meyer2008MNRAS.391.1712M}; but see \citealt{Springob+2005}). The exact correction factor depends on the definition of line width (e.g., \citealt{Rhee2005JASS...22...89R}; \citealt{Meyer2008MNRAS.391.1712M}). 

\begin{figure}
\epsscale{1.1}
\plotone{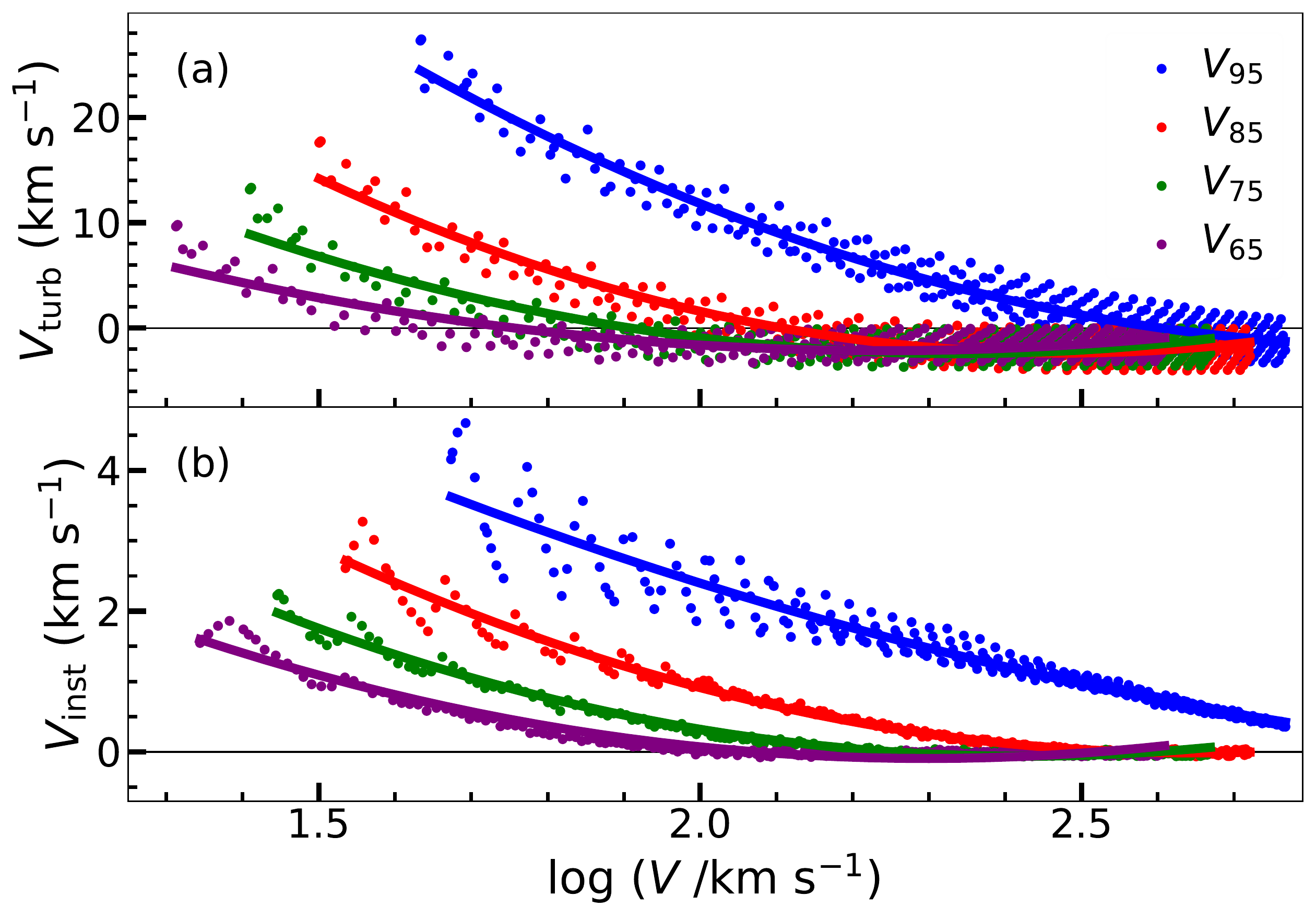}
\caption{Line width correction terms for (a) turbulence motions and (b) instrumental resolution, for lines of different velocity width ($V$). The points with different colors represent simulation results for the effect of broadening on the line widths ($V_{95}$, $V_{85}$, $V_{75}$, and $V_{65}$). The corresponding curves are second-order polynomial fits listed in Table~3, which assume a full width at half maximum instrumental resolution of $v_{\rm inst} = 11$ \kms.} 
\label{fig:lwcor}
\end{figure}

We use mock spectra to investigate the proper correction terms for our newly defined line widths. We generate double-horn profiles\footnote{If the line width is small, as in low-mass galaxies, the broadening effects of a double-horn profile and a single Gaussian profile are similar.} with total line width 12--600 \kms, which cover the extremes of the range of line widths observed in galaxies (e.g., \citealt{Haynes+2011AJ....142..170H}), and then convolve each mock spectrum with two Gaussian functions to simulate the smearing effects caused by instrumental resolution and turbulence motions.  We assume that the turbulence motions of \HI\ disks have a typical velocity dispersion of $\sigma_{\rm turb} = 10$ \kms\ \citep{Lelli2016AJ....152..157L}.  The instrumental resolution differs from observation to observation (Table~\ref{tab:basic}). For illustrative purposes, we highlight the fiducial case of an observation with a full width at half maximum instrumental resolution of $v_{\rm inst} = 11$ \kms\ (equivalent to $\sigma_{\rm inst} = 4.7$ \kms).  The channel width is set to  5.5 \kms, mimicking data from the ALFALFA survey.  Comparing the line widths before and after convolution, we quantify the broadening correction factors $V_{\rm turb}$ and $V_{\rm inst}$ as a function of line width $V$. Figure~\ref{fig:lwcor} shows the simulation results, and Table~\ref{tab:SimuFit} lists the coefficients of the second-order polynomial fits used to compute the broadening corrections.
 
\begin{deluxetable}{lccccc}
\tabletypesize{\footnotesize}
\tablecolumns{6}
\tablecaption{Fitting Parameters for Line Width Correction}
\tablehead{
\colhead{Broadening} &
\colhead{Line Width} &
\colhead{$a$} &
\colhead{$b$} &
\colhead{$c$} 
 \\ 
\colhead{(1)} &
\colhead{(2)} &
\colhead{(3)} &
\colhead{(4)} &
\colhead{(5)} 
}
\startdata
\multirow{4}{*}{$V_{\rm turb}$}  & $V_{95}$ & 15.38  & $-$90.35 & 130.99 \\
                                 & $V_{85}$ & 17.30  & $-$85.76 & 103.89 \\ 
                                 & $V_{75}$ & 13.08  & $-$61.20 & "69.16 \\ 
                                 & $V_{65}$ & "9.14  & $-$40.97 & "43.77 \\ 
\hline                                         
\multirow{4}{*}{$V_{\rm inst}$}  & $V_{95}$ & 1.08   & "$-$7.72 & "13.51 \\
                                 & $V_{85}$ & 2.20   & $-$11.65 & "15.42 \\ 
                                 & $V_{75}$ & 2.13   & $-$10.31 & "12.43 \\ 
                                 & $V_{65}$ & 1.86   & "$-$8.55 & ""9.73 \\ 
\enddata
\tablecomments{Col. (1): Broadening effect. Col. (2) Line width. Col. (3)--(5): Fitting parameters for the polynomial $y = a x^2 + b x +c$, where $x$ is the logarithm of the line width, and $V_{\rm inst}$ is derived by assuming a full width at half maximum instrumental resolution of $v_{\rm inst} = 11$ \kms.}
\label{tab:SimuFit}
\end{deluxetable}

\subsubsection{Correction for Inclination Angle}
\label{subsec:LinesCorIncl}

Although the majority of our galaxies have robust inclination angles for the \HI\ disk derived from spatially resolved kinematics, we choose not to use them and instead estimate the inclination angle from the axial ratio derived from optical images of the stellar light distribution.  We adopt this procedure with the aim of applying it to large existing and future single-dish \HI\ surveys, which, lacking spatial resolution, invariably must avail to optical or near-infrared images to estimate galaxy inclinations.  An implicit key assumption is that the stars and \HI\ gas are coplanar.  Following \cite{Hubble1926ApJ....64..321H}, we estimate the inclination angle $i$ as

\begin{equation}
{\rm cos}^2 \,i = \frac{q^2-q^2_0}{1-q^2_0},
\label{equ:incl}
\end{equation}

\noindent
where the axial ratio $q = b/a$, with $a$ the semi-major and $b$ the semi-minor axis of the isophote at a $B$-band surface brightness level of $\mu = 25$ mag~arcsec$^{-2}$.  We use values of $q$ from HyperLeda (\citealt{Paturel2003_hyperledaII}; \citealt{Makarov2014A&A...570A..13M})\footnote{{\url http://leda.univ-lyon1.fr}}.  The intrinsic disk thickness $q_0$ is not constant among galaxies but depends on the morphological type (\citealt{Sandage1970ApJ...160..831S}; \citealt{Aaronson1980ApJ...237..655A}; \citealt{Fouque1990ApJ...349....1F}), luminosity (\citealt{Rodrguez2013MNRAS.434.2153R}; \citealt{Roychowdhury2013MNRAS.436L.104R}), and stellar mass (\citealt{Sanchez-Janssen2010MNRAS.406L..65S}).  

Bearing in mind future applications, we prefer to estimate $q_0$ from the stellar mass \citep{Sanchez-Janssen2010MNRAS.406L..65S}, which generally is or will become available for most large galaxy surveys.  Luminosity is less ideal because galaxies span a large range of stellar populations and hence mass-to-light ratios.  In our opinion, morphological types are least desirable because they are highly subjective, strongly sensitive to image resolution and wavelength, and, in any case, often unavailable for very large surveys.  Most of our galaxies have integrated $K_s$-band photometry from the Two Micron All Sky Survey \citep{Jarrett2003AJ....125..525J, Skrutskie2006AJ....131.1163S, Fingerhut2010ApJ...716..792F}, which we convert to stellar mass assuming a stellar mass-to-light ratio of 0.6 \msun/$L_{\odot}$ \citep{McGaugh2014AJ....148...77M}. The $K_s$ band minimizes stellar population differences between early-type galaxies and late-type galaxies, and the variation in $(M/L)_K$ is about 0.2 dex for different galaxy types (Bell et al. 2003). More sophisticated prescriptions can be entertained \citep{KormendyHo2013}, but a constant mass-to-light ratio suffices for the current purposes.  The stellar masses for the minority of sources lacking $K_s$-band data were taken from the literature (\citealt{Sheth2010PASP..122.1397S}; \citealt{McGaugh2012AJ....143...40M}) or calculated from $R$-band photometry and a stellar mass-to-light ratio estimated from $B-R$ colors (\citealt{Swaters2002A&A...390..863S}; \citealt{Bell2003ApJS..149..289B}).  We adopt absolute magnitudes of the Sun from \citet{Willmer2018ApJS..236...47W}.  

\begin{figure}
\epsscale{1.15}
\plotone{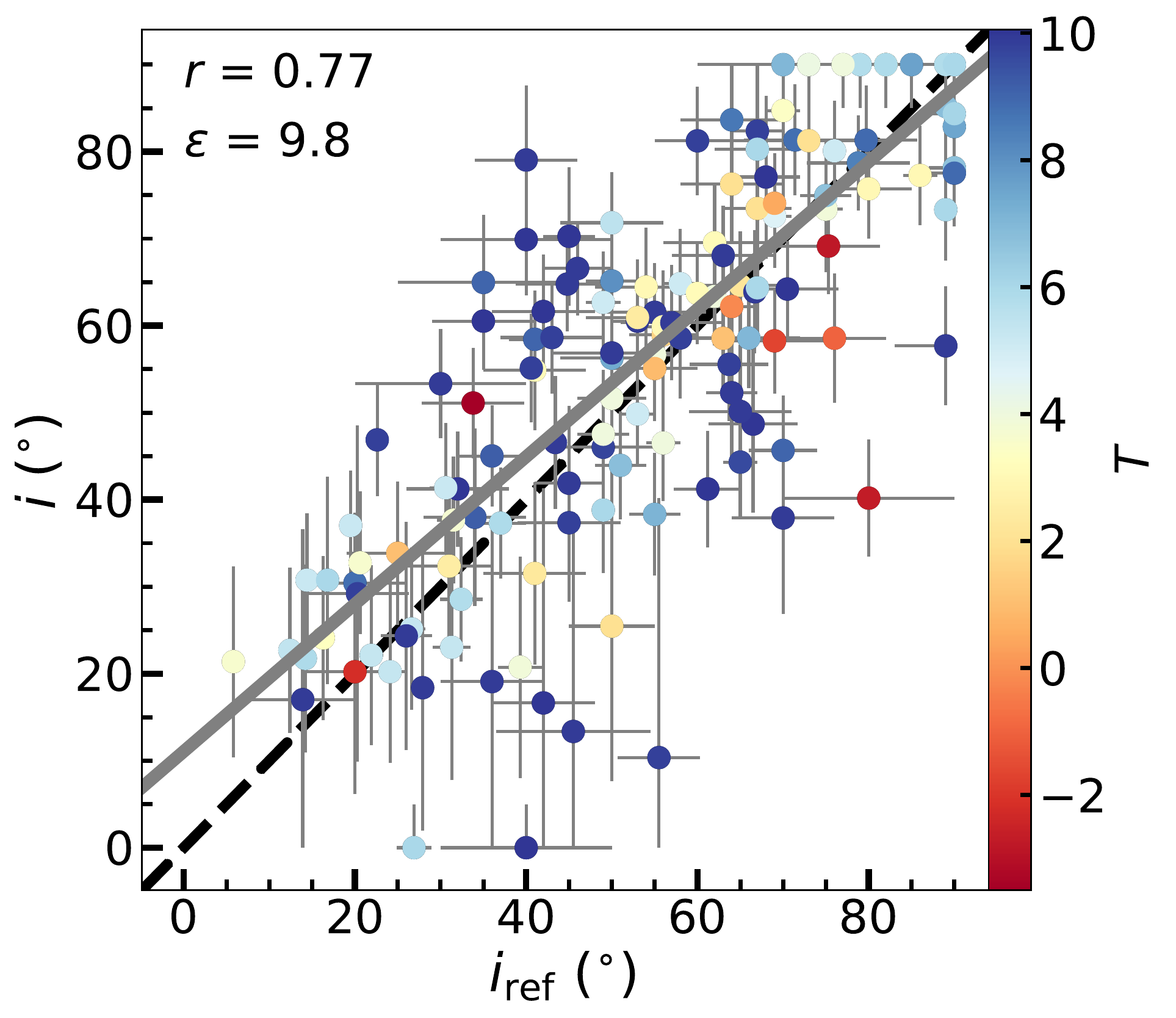}
\caption{Comparison of inclination angles from the literature ($i_{\rm ref}$) with those derived in this work ($i$), color-coded by morphological type. The black dashed line shows the 1:1 relation, and the grey solid line is the best-fit linear regression.  The upper-left corner gives the Pearson correlation coefficient $r$ and the scatter $\epsilon$ of the best-fit linear regression.}
\label{fig:inclComp}
\end{figure}

The final inclination angles ($i$), listed in Table~1, consider the uncertainties in $q$ from HyperLeda, a typical uncertainty of 0.05 for $q_0$ from \cite{Sanchez-Janssen2010MNRAS.406L..65S}, and an additional systematical uncertainty of 5$^{\circ}$ for $i$ due to potential photometric misalignment between the \HI\ and stellar disks \citep{Barnes2003AJ....125.1164B, Serra2012MNRAS.422.1835S}.  Figure \ref{fig:inclComp} compares our inclination angles with those from the literature ($i_{\rm ref}$) derived from tilted-ring fits of the \HI\ velocity field (\citealt{Verheijen2001ApJ...563..694V}; \citealt{Noordermeer2007MNRAS.376.1513N}; \citealt{Martinsson2013AA...557A.131M};  \citealt{Kamphuis2015MNRAS.452.3139K}; \citealt{Oh2015AJ....149..180O}; \citealt{Lelli2016AJ....152..157L}; \citealt{Wang2017MNRAS.472.3029W}), from the axial ratio of the \HI\ disk (\citealt{Begum2008MNRAS.386.1667B}; \citealt{deBlok2008AJ....136.2648D}; \citealt{Koribalski2018MNRAS.478.1611K}), and from the velocity field of stars (\citealt{Cappellar2013MNRAS.432.1709C}).  The two sets of values are strongly correlated with a Pearson correlation coefficient of $r=0.77$.  We derive a scatter of $\epsilon = 9.8^{\circ}$ using the {\tt LinMix} package \citep{Kelly2007LinMix}, which uses Bayesian inferences and Monte Carlo Markov Chain sampling to perform linear regression while accounting for uncertainties in both variables.  The scatter increases toward low inclination angles, which can be especially uncertain even when derived from kinematical methods \citep{Kamphuis2015MNRAS.452.3139K}.
In the subsequent analysis, we exclude four galaxies with $i < 10^{\circ}$ to avoid extremely large and uncertain inclination angle corrections.

We warn that the stars and \HI\ may be misaligned in early-type (E and S0, or $T\leq0$) galaxies \citep{Morganti2006}. Within the small-number statistics of our sample, we see no strong evidence of misalignment between stars and \HI\ in early-type galaxies compared to late-type galaxies (Figure \ref{fig:inclComp}).

\subsubsection{Final Corrections}
\label{subsec:LinesCorFinal}

In summary, the final, corrected line widths (Table~\ref{tab:msHI}) can be expressed as

\begin{equation}
V_{jc} = {{(V_{j} - V_{\rm inst})/(1+z) - V_{\rm turb}}\over{2 \ {\rm sin}\ i}}.
\end{equation}

\noindent
where $j = \{95, 85, 75, 65\}$ and the corresponding values for $V_{\rm inst}$ and $V_{\rm turb}$ are given in Table~\ref{tab:SimuFit}.  The estimated errors do not consider uncertainties propagated from $z$, $V_{\rm inst}$, or $V_{\rm turb}$, which are negligible compared with the uncertainties due to $i$.  In a few rare cases the corrected line widths may fall below the minimum value set by $k \sqrt{\sigma^2_{\rm inst}+\sigma_{\rm turb}^2(1+z)^2}$, where the factor $k = 3.84, 2.78, 2.25$, and 1.99 for $V_{95}$, $V_{85}$, $V_{75}$, and $V_{65}$, respectively.  For these we set the corrected line width to this minimum value and consider it an upper limit.

\subsubsection{Relating Line Widths to Rotation Velocities}

\begin{figure}
\epsscale{1.2}
\plotone{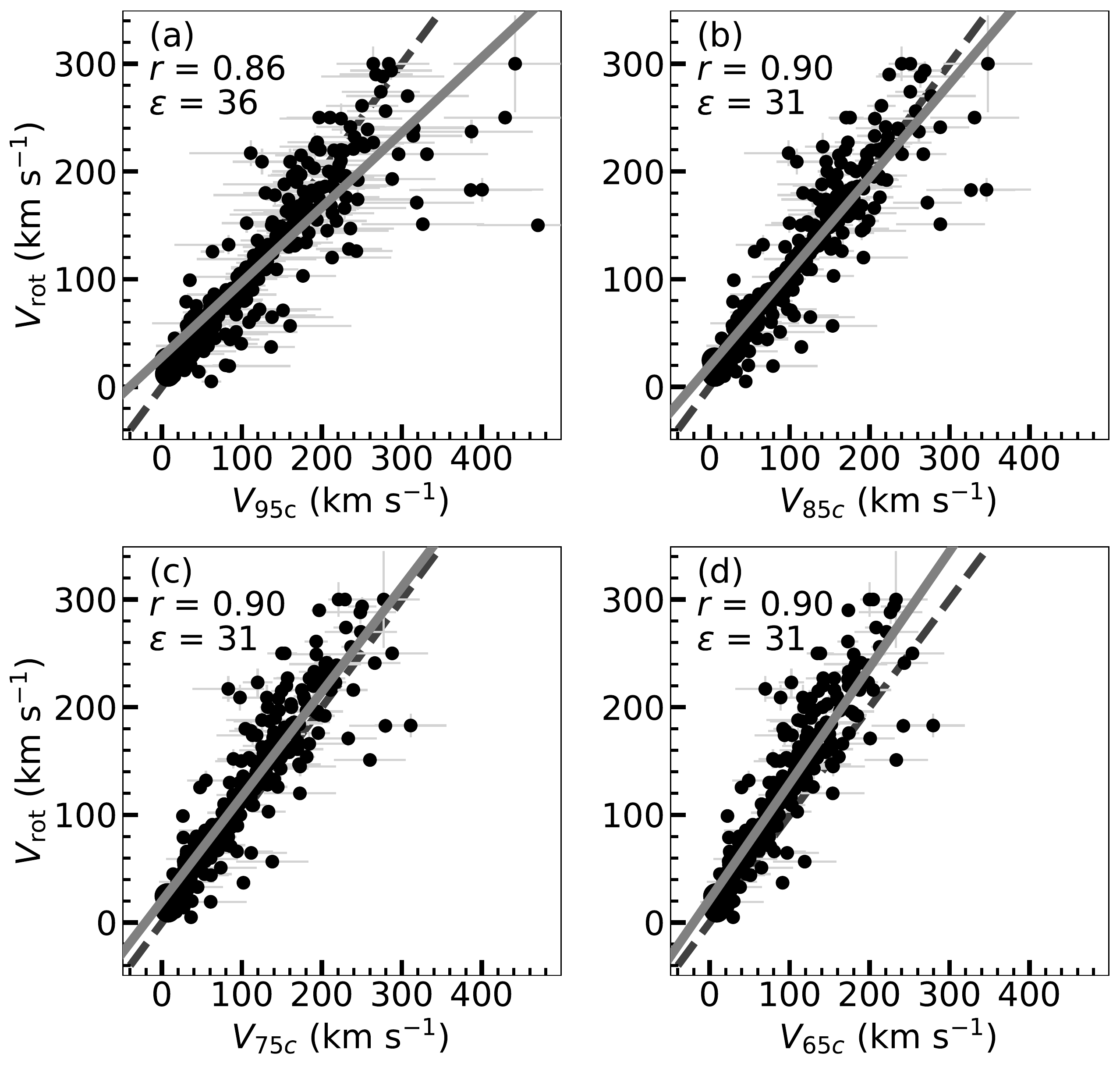} 
\caption{Linear relations between \vrot\ and corrected line width (a) $V_{95c}$, (b) $V_{85c}$, (c) $V_{75c}$, and (d) $V_{65c}$. The black dashed line shows the 1:1 relation, and the grey solid line gives the best-fit linear regression.  Each panel gives the Pearson correlation coefficient $r$ and the scatter $\epsilon$ of the regression fit.}
\label{fig:fraction}
\end{figure}

Which, among the four newly defined measures of line width, best predicts the true rotation velocity? Figure~\ref{fig:fraction} shows the correlation between $V_{\rm rot}$ and the measured line widths, corrected for redshift, instrumental resolution, turbulence, and inclination angle as described in Section 4.1.3.  All four measures of line width are strongly correlated with $V_{\rm rot}$, with Pearson correlation coefficient $r \approx 0.9$.  A regression analysis using {\tt LinMix} yields nearly linear slopes and a scatter of $\sim 30$ \kms.  We adopt $V_{85c}$ as the final estimator for rotation velocity because it exhibits less scatter than $V_{95c}$ and because it gives a closer one-to-one match with $V_{\rm rot}$ than does $V_{75c}$ or $V_{65c}$.

\begin{figure}
\epsscale{1.1}
\plotone{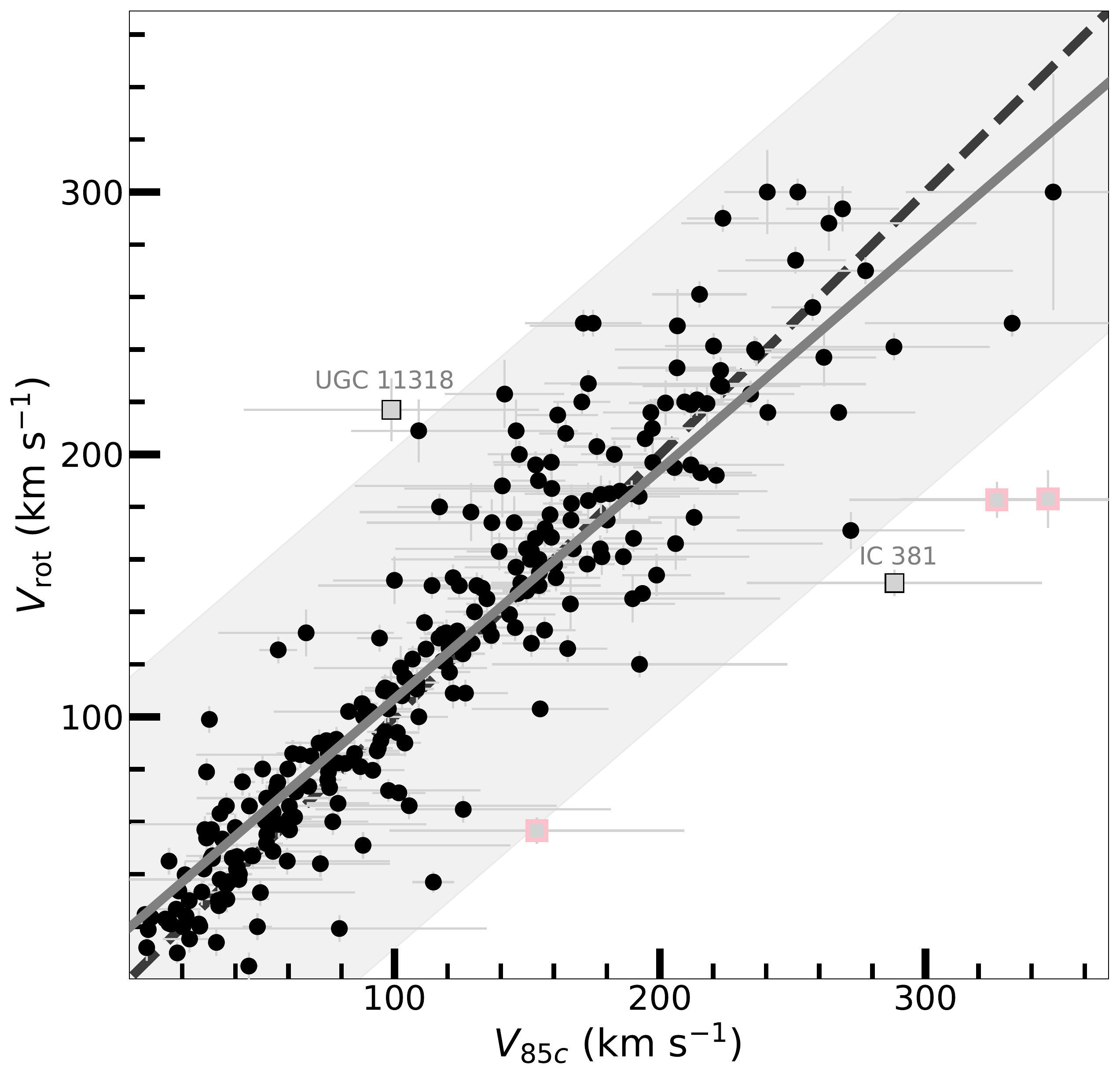}
\caption{Correlation between \vrot\ and $V_{85c}$ for the galaxies in our calibration sample. The black dashed line shows the 1:1 relation, and the grey solid line shows the best-fit linear relation, with the grey shaded region denoting the $3\,\epsilon$ scatter of the fit. The grey squares are galaxies that deviate by more than $3\,\epsilon$; grey squares highlighted in pink are galaxies for which our adopted inclination angle differs from that used in the literature.}
\label{fig:outliers}
\end{figure}

Closer inspection of the correlation between $V_{\rm rot}$ and $V_{85c}$ reveals five outliers beyond the $3\, \epsilon$ scatter (Figure~\ref{fig:outliers}). Three of them are due to large differences between our inclination angles and those adopted in the literature for deriving $V_{\rm rot}$. UGC~11318 has its value of $V_{85c}$ overestimated because of confusion with neighboring sources in its single-dish spectrum (Figure~\ref{fig:ctm}). The global profiles of IC~381 from single-dish and interferometric observations are consistent, but its literature values of $V_{\rm rot}$ are wildly discrepant (151~\kms: \citealt{Rhee2005JASS...22...89R}; 304 ~\kms: HyperLeda). After excluding these five outliers, the relation between rotational velocity and line width is

\begin{equation}
V_{\rm rot} = (0.94\pm 0.02)\, V_{85c}+(13.33\pm 3.31),
\label{equ:fitdrp}
\end{equation}

\noindent
and the scatter is $\epsilon = 27$ \kms.

\begin{figure}
\epsscale{1.15}
\plotone{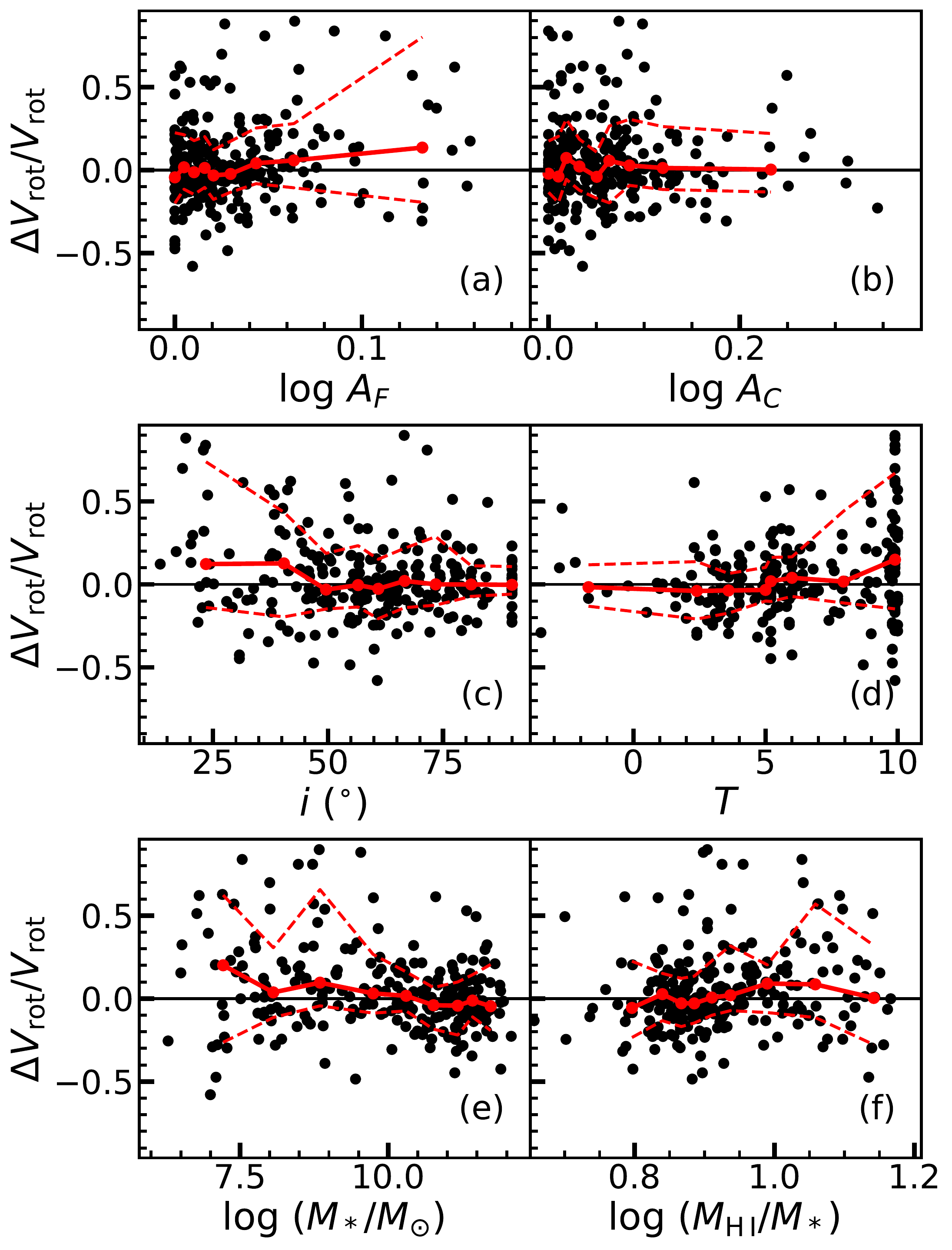}
\caption{Residuals ($\Delta V_{\rm rot}/V_{\rm rot}$) from the \vrot$-V_{85c}$ relation as a function of (a) flux asymmetry $A_F$, (b) concentration asymmetry $A_C$, (c) inclination angle $i$, (d) morphological type index $T$, (e) stellar mass $M_*$, and (f) ratio of \HI\ mass to stellar mass. The black horizontal line denotes zero offset. The solid red curves show the median value, and the dashed red curves show the 16th and 84th percentiles of the distribution.}
\label{fig:secondary}
\end{figure}

Defining $\Delta V_{\rm rot}$ as the difference between the true rotation velocity and the rotation velocity predicted from $V_{85c}$ (Equation~\ref{equ:fitdrp}), Figure~\ref{fig:secondary} investigates possible secondary dependences of $\Delta V_{\rm rot}/V_{\rm rot}$ on profile asymmetry ($A_F$, $A_C$), inclination angle ($i$), morphological type ($T$), stellar mass ($M_*$), and gas-to-stellar mass ratio (\MHI/$M_*$).  The rotation velocity is slightly overestimated and the scatter in $\Delta V_{\rm rot}/V_{\rm rot}$ generally increases toward galaxies with large flux asymmetry.  The same holds for galaxies of low inclination, low mass, high gas fraction, and late type ($T = 10$ are dwarf irregular galaxies), precisely in the very systems wherein $q_0$, and hence $i$, is poorly known.

\begin{deluxetable*}{cD@{$\pm$}DD@{$\pm$}DD@{$\pm$}DD@{$\pm$}DD@{$\pm$}DD@{$\pm$}DD@{$\pm$}DD@{$\pm$}DD@{$\pm$}DDD@{$\pm$}DD@{$\pm$}Dc}
\centering
\small\addtolength{\tabcolsep}{-3.5pt}
\tabletypesize{\footnotesize}
\tablecolumns{14}
\tablewidth{0pt} 
\tablecaption{Physical Parameters Derived from the \HI\ Spectra}
\tablehead{
\colhead{Galaxy} &
\multicolumn4c{$V_c$} &
\multicolumn4c{$F$} &
\multicolumn4c{$V_{95c}$} &
\multicolumn4c{$V_{85c}$} &
\multicolumn4c{$V_{75c}$} &
\multicolumn4c{$V_{65c}$} &
\multicolumn4c{$A_F$} &
\multicolumn4c{$A_C$} &
\multicolumn4c{$C_V$} &
\multicolumn2c{SNR} &
\multicolumn4c{log \MHI} &
\multicolumn4c{log \mdyn}  &
\colhead{Notes} \\ 
\colhead{} &      
\multicolumn4c{(\kms)} & 
\multicolumn4c{(Jy\ \kms)} &
\multicolumn4c{(\kms)} &
\multicolumn4c{(\kms)} &
\multicolumn4c{(\kms)} &
\multicolumn4c{(\kms)} &
\multicolumn4c{} &
\multicolumn4c{} & 
\multicolumn4c{} &  
\multicolumn2c{} & 
\multicolumn4c{(\msun)} &  
\multicolumn4c{(\msun)} &
\colhead{}
\\
\colhead{(1)} &
\multicolumn4c{(2)} &
\multicolumn4c{(3)} &
\multicolumn4c{(4)} &
\multicolumn4c{(5)} &
\multicolumn4c{(6)} &
\multicolumn4c{(7)} &
\multicolumn4c{(8)} &
\multicolumn4c{(9)} &
\multicolumn4c{(10)} &
\multicolumn2c{(11)} &
\multicolumn4c{(12)} &
\multicolumn4c{(13)}&
\colhead{(14)}  
}
\decimals 
\startdata
UGC~1281 & 156 & 2 & 38.98 & 6.16 & 55 & 10 & 51 & 3 & 46 & 2 & 40 & 2 & 1.03 & 0.07 & 1.06 & 0.11 & 3.44 & 0.34 & 578.0 & 8.44 & 0.13 & 9.63 & 0.14 & 1, 2 \\
UGC~2023 & 602 & 2 & 17.23 & 2.73 & 64 & 76 & 56 & 55 & 50 & 45 & 44 & 39 & 1.00 & 0.07 & 1.03 & 0.10 & 4.31 & 0.43 & 351.9 & 8.62 & 0.13 & 9.77 & 0.87 & 2 \\
UGC~2034 & 578 & 2 & 30.19 & 4.77 & 38 & 15 & 31 & 9 & 26 & 8 & 22 & 6 & 1.00 & 0.07 & 1.01 & 0.10 & 4.47 & 0.45 & 385.1 & 8.86 & 0.13 & 9.52 & 0.31 & 2 \\
UGC~2053 & 1024 & 2 & 16.00 & 2.53 & 34 & 8 & 30 & 3 & 26 & 2 & 22 & 2 & 1.02 & 0.07 & 1.09 & 0.11 & 4.26 & 0.43 & 211.2 & 8.72 & 0.13 & 9.42 & 0.16 & 2 \\
UGC~3371 & 814 & 2 & 25.21 & 3.99 & 91 & 24 & 84 & 17 & 76 & 15 & 68 & 13 & 1.01 & 0.07 & 1.10 & 0.11 & 3.00 & 0.30 & 339.4 & 8.99 & 0.13 & 10.25 & 0.22 & 2 \\
UGC~3817 & 437 & 2 & 10.81 & 1.71 & 15 & 5 & 15 & 1 & 13 & 1 & 12 & 1 & 1.04 & 0.07 & 1.11 & 0.11 & 3.97 & 0.40 & 167.7 & 8.29 & 0.13 & 8.84 & 0.17 & 2 \\
UGC~4173 & 859 & 2 & 25.59 & 4.05 & 35 & 7 & 31 & 2 & 27 & 1 & 23 & 1 & 1.05 & 0.07 & 1.14 & 0.11 & 3.87 & 0.39 & 830.3 & 9.23 & 0.13 & 9.70 & 0.15 & 2 \\
UGC~4278 & 559 & 2 & 43.86 & 6.94 & 87 & 15 & 78 & 4 & 70 & 3 & 63 & 3 & 1.03 & 0.07 & 1.14 & 0.11 & 2.80 & 0.28 & 102.3 & 9.06 & 0.13 & 10.23 & 0.14 &  \\
UGC~4325 & 506 & 2 & 20.50 & 3.24 & 91 & 17 & 74 & 6 & 63 & 4 & 53 & 4 & 1.05 & 0.08 & 1.30 & 0.13 & 3.48 & 0.35 & 34.4 & 8.69 & 0.13 & 10.01 & 0.15 &  \\
UGC~4499 & 689 & 2 & 26.31 & 4.16 & 60 & 11 & 55 & 4 & 49 & 3 & 43 & 2 & 1.04 & 0.07 & 1.07 & 0.11 & 3.20 & 0.32 & 539.1 & 9.02 & 0.13 & 9.97 & 0.15 & 2 \\
\enddata
\tablecomments{Col. (1): Galaxy name. Col. (2): Flux intensity-weighted central velocity. Col. (3): Total integrated flux of \HI\ line. Cols. (4)--(7): Corrected line width measured at 95\%, 85\%, 75\%, and 65\% of the total flux. Col. (8): Corrected flux asymmetry. Col. (9): Corrected concentration asymmetry. Col. (10): Corrected concentration of \HI\ profile. Col. (11): SNR of the profile. Col. (12): \HI\ mass; if we assume an uncertainty in the distance of 10\% and a flux uncertainty of 15\%, the typical uncertainty of $\log$\MHI\ is 0.13 dex. Col. (13): Dynamical mass determined using Equation \ref{equ:mdynour}; the typical uncertainty is 0.23 dex, taking into consideration the uncertainties of \MHI, $V_{85c}$, and $i$. Col. (14): 1 = mask generated for the spectrum; 2 = baseline subtracted in this work.  (Table \ref{tab:msHI} is published in its entirety in machine-readable format. A portion is shown here for guidance regarding its form and content.)}
\label{tab:msHI}
\end{deluxetable*}

\subsection{Dynamical Masses and Halo Masses}
\label{subsec:mdyn}

For a rotationally supported \HI\ disk, the dynamical mass 

\begin{equation}
M_{\rm dyn} = \frac{V_{\rm rot}^2 R_{\rm H\, \textsc{i}}}{G} = 2.31\times10^5 \,M_\odot \left(\frac{V_{\rm rot}^2}{\mathrm{km\ s^{-1}}}\right) \left(\frac{R_{\rm H\, \textsc{i}}}{\mathrm{kpc}}\right),
\label{equ:mdyns}
\end{equation}

\noindent
where $R_{\rm H\, \textsc{i}}$ is the \HI\ radius.   Single-dish \HI\ spectra generally furnish no direct information on $R_{\rm H\, \textsc{i}}$.  We take advantage of the fact that normal, isolated, gas-rich galaxies obey a tight empirical relation between $R_{\rm H\, \textsc{i}}$ and \HI\ mass (\citealt{BroeilsRhee1997}; \citealt{VerheijenSancisi2001UM_HI}; \citealt{Swaters2002AA...390..829S}; \citealt{Noordermee2005whisp(sa)}; \citealt{Begum2008MNRAS.386.1667B}; \citealt{Wang2013MNRAS.433..270W, Wang2014MNRAS.441.2159W}; \citealt{Martinsson2016AA...585A..99M}).  We adopt the calibration of \cite{Wang+2016HI-size-mass}:

\begin{equation}
\log \left(\frac{R_{\rm H\, \textsc{i}}}{{\rm kpc}}\right) = (0.51\pm0.00) \log \left(\frac{M_{\rm H\, \textsc{i}}}{M_\odot}\right) - (3.59\pm0.01).
\label{equ:mdyns}
\end{equation}

\noindent
Remarkably, this relation, which has an rms scatter of only $\sim 0.06$ dex, holds over 5 orders of magnitude in \HI\ mass ($M_{\rm H\, \textsc{i}} \approx 10^{5.5}-10^{10.5}\,M_\odot$) and nearly 3 orders of magnitude in size ($R_{\rm H\, \textsc{i}} \approx 0.15 - 110$ kpc).
Combining Equations~\ref{equ:fitdrp}--\ref{equ:mdyns}, the dynamical mass within \RHI\ can be estimated purely from observable quantities:

\begin{equation}
\log\ M_{\rm dyn}  = 2\log\ (0.94V_{85c}+13.33)+ 0.51\log\ (F\,D_L^2)+4.51,
\label{equ:mdynour}
\end{equation}

\noindent 
with $M_{\rm dyn}$ in units of $M_\odot$, $V_{85c}$ in \kms, $F$ in Jy~\kms, and $D_L$ in Mpc.

\begin{figure}[ht!]
\epsscale{1.1}
\plotone{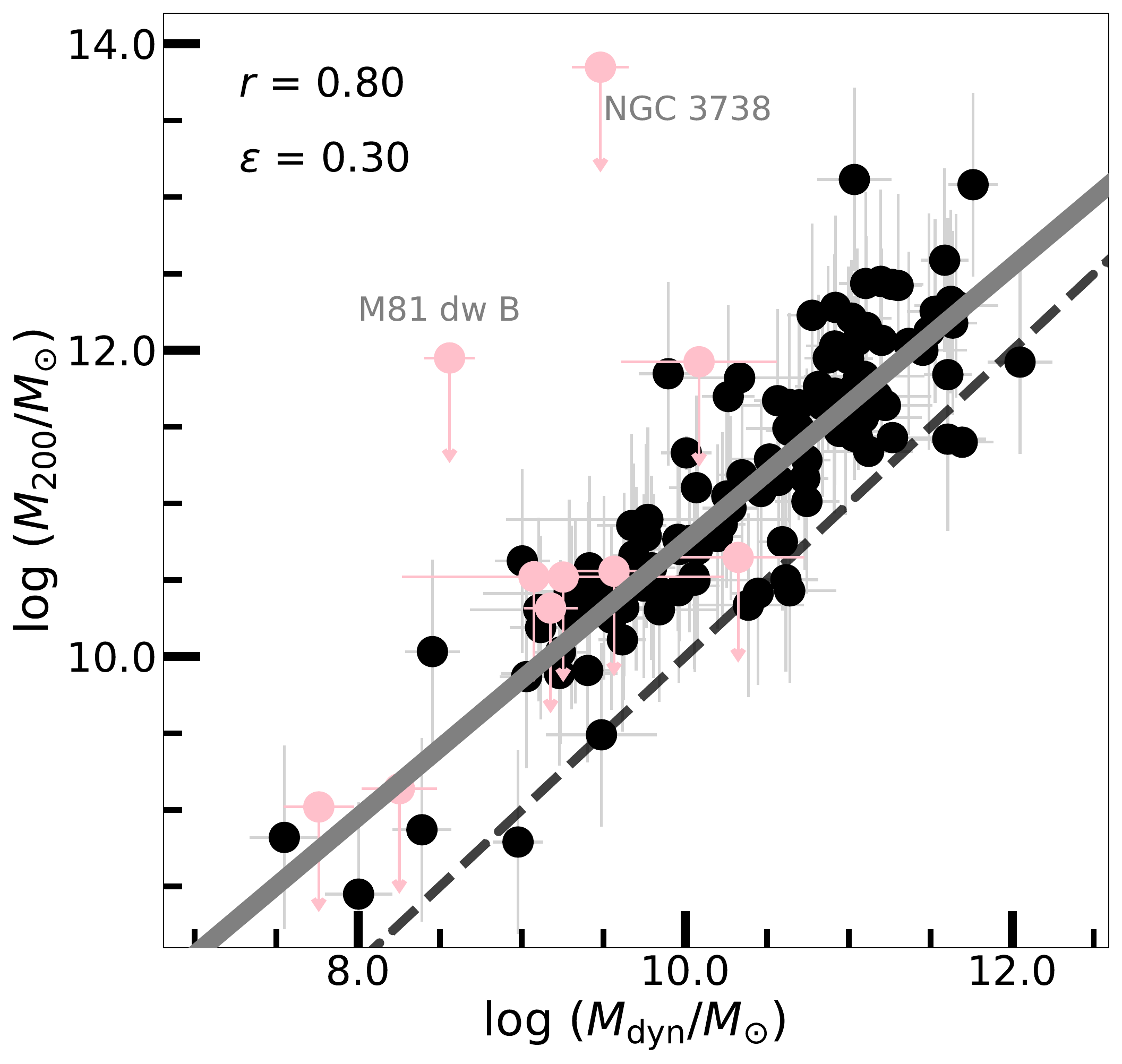}
\caption{Relation between \mtwo\ and \mdyn. The grey line shows the best fit, and the black dashed line gives the 1:1 relation. Pearson correlation coefficient $r$ and the scatter $\epsilon$ are shown in the upper-left corner. Galaxies with upper limits in \mtwo\ are plotted as pink points.}
\label{fig:mdyn}
\end{figure}

The dynamical mass of a galaxy on the scale of the \HI\ disk is closely related to but still falls far short of the mass of the dark matter-dominated halo. The mass of the halo is commonly defined as the mass (\mtwo) within $r_{200}$, the radius within which the average density is 200 times the critical density of the Universe (e.g., \citealt{Navarro1996ApJ...462..563N}; \citealt{deBlok2008AJ....136.2648D}; \citealt{Oh2011AJ....142...24O}; \citealt{Li2019MNRAS.482.5106L}).  Nearly half (120/265) of the galaxies in our sample have published values of \mtwo\ (Table \ref{tab:basic}).  Apart from the handful of dwarf galaxies studied by \citet{Forbes2018MNRAS.481.5592F}, who used inclination-corrected peak rotation velocities to derive halo masses, most of the halo masses were derived from detailed models of their rotation curves.  The correlation between \mtwo\ and \mdyn, shown in Figure~\ref{fig:mdyn}, has a Pearson correlation coefficient $r = 0.80$. The best-fit linear relation, taking into account upper limits in \mtwo, has a scatter of $\epsilon = 0.30$ dex and the form

\begin{equation}
\begin{split}
\log\ & M_{200} = (0.90\pm 0.07)\log\ M_{\rm dyn} +(1.76\pm 0.73)\\
 &              = 1.80\log\ (0.94V_{85c}+13.33)+0.46\log\ (F\,D_L^2)+5.82.
\end{split}
\label{equ:m200fit}
\end{equation}

There are no obvious secondary dependences between the scatter in Figure~\ref{fig:mdyn} and other parameters, including profile asymmetry,  inclination angle, morphological type, stellar mass, or gas-to-stellar mass ratio.

\section{Summary}
\label{sec:sum}

We develop a new method based on the concept of the curve-of-growth (CoG) and apply it to the integrated \HI\ spectra to derive a set of robust observational parameters that can be related to the physical properties of galaxies.  Apart from the trivial quantities of central velocity ($V_c$) and total line flux ($F$), our method can be used to measure a new set of line widths ($V_{65}$, $V_{75}$, $V_{85}$, $V_{95}$) that enclose different percentages of the total line flux.  We further use the properties of the CoG to introduce two asymmetry parameters that characterize differences in the integrated flux ($A_F$) and flux distribution ($A_C$) of the red and blue sides of the line, as well as the parameter $C_V$, which specifies the degree of concentration of the overall profile.  We generate comprehensive sets of mock spectra to assess the performance of our procedure and to derive realistic systematic uncertainties for our proposed parameters.  We also offer customized prescriptions for correcting the new line widths for the effects of instrumental resolution and turbulence broadening. 

Utilizing a sample of 265 nearby galaxies with available spatially resolved kinematics, we investigate the feasibility of using the newly defined line widths to predict the rotational velocity ($V_{\rm rot}$) of the galaxy.  The inclination angle-corrected line width $V_{85c}$ emerges as the most promising estimator of $V_{\rm rot}$; the rotation velocity can be recovered within a typical rms scatter of 27 \kms.  Taking advantage of the empirical relation between \HI\ radius and \HI\ mass from Wang et al. (2016), we offer a new formalism to estimate the dynamical mass (\mdyn) within the \HI-emitting radius of gas-rich galaxies. Our formalism, which yields dynamical masses to an accuracy of $\sim 0.3$ dex, is based solely on quantities that can be derived efficiently and robustly from integrated \HI\ spectra and can be a powerful tool for current and future large-scale extragalactic \HI\ surveys.  We further extend the dynamical mass calibration to the scale of the dark matter halo ($r_{200}$) with the aid of a subset of galaxies having published halo masses (\mtwo).

\acknowledgments
This work was supported by the National Science Foundation of China (11721303, 11991052) and the National Key R\&D Program of China (2016YFA0400702).  We thank Sandra Faber, Linhua Jiang, Se-Heon Oh, D. J. Pisano, Jinyi Shangguan, M. A. W. Verheijen, Bitao Wang, Shun Wang, T. Westmeier, and Pei Zuo for useful advice and discussions.  This research has made use of the NASA/IPAC Extragalactic Database, which is funded by the National Aeronautics and Space Administration and operated by the California Institute of Technology. We acknowledge the usage of the HyperLeda database ({\url http://leda.univ-lyon1.fr}). This publication uses data products from the Two Micron All Sky Survey, which is a joint project of the University of Massachusetts and the Infrared Processing and Analysis Center/California Institute of Technology, funded by the National Aeronautics and Space Administration and the National Science Foundation. We used Astropy, a community-developed core Python package for astronomy \citet{AstropyCollaboration2013A&A...558A..33A}.  

\newpage

\appendix
\section{Notes on Special Cases}
\label{appsec:HIp}
\begin{figure*}
\centering
\figurenum{A1}
\subfigure{\includegraphics[width=1.0\textwidth]{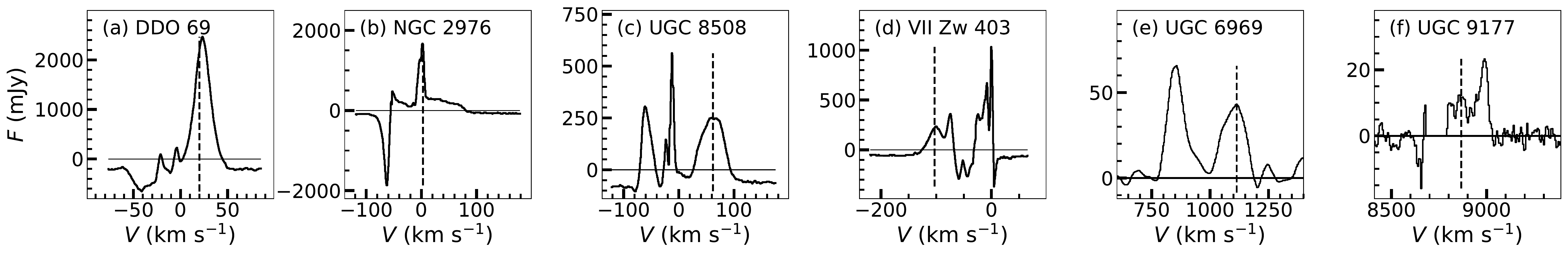}}
\caption{Discarded single-dish \HI\ spectra. The optical central velocities 
are shown as black dashed lines.}
\label{fig:HIspecDis}
\end{figure*}

Four spectra from \citet{TifftCocke1988ApJS...67....1T} show suspicious-looking profiles, as shown in Figures~\ref{fig:HIspecDis}a--\ref{fig:HIspecDis}d. The high flux densities and the highly peculiar profiles strongly suggest that these spectra are contaminated by artifacts.  We exclude them from our sample.  The optical central velocity of UGC~6969 is centered on the right peak of the spectrum spectrum from \citet{Richter1991AAS...87..425R}.  The spectrum of UGC~9177 from \citet{Haynes+2011AJ....142..170H} seems to be incomplete, because there is an obvious mask to the blue side of the signal, and the optical central velocity is not at the center of the \HI\ emission. We also discarded these two spectra.  

\begin{figure*}
\centering
\figurenum{A2}
\subfigure{\includegraphics[width=1.0\textwidth]{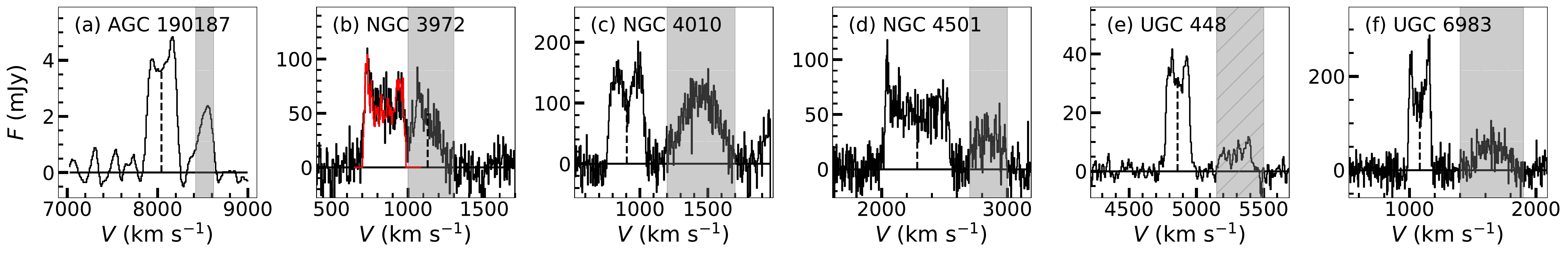}}
\caption{Single-dish \HI\ spectra that are partly masked (grey shaded regions). The black dashed line in each panel is the central velocity of the single-dish spectrum from the literature. The red profile in panel (b) shows the interferometric observations
of NGC~3972.}
\label{fig:HIspecMask}
\end{figure*}

Another six spectra show possible nearby companions (Figure~\ref{fig:HIspecMask}). The central velocities from the literature (AGC~190187: \citealt{Mould1993ApJ...409...14M}; NGC~3972, NGC~4010, NGC~4501 and UGC~6983: \citealt{Springob+2005}; UGC~448: \citealt{Haynes+2011AJ....142..170H}) are shown as black dashed lines.  We manually masked the signal that appears to arise from a contaminating source (grey shaded region).  Determination of the mask for the spectrum of NGC~3972 was aided by comparison with the interferometric observation of Verheijen \& Sancisi (2001; part of profile highlighted in red). 
 
\begin{figure*}
\centering
\figurenum{A3}
\subfigure{\includegraphics[width=0.35\textwidth]{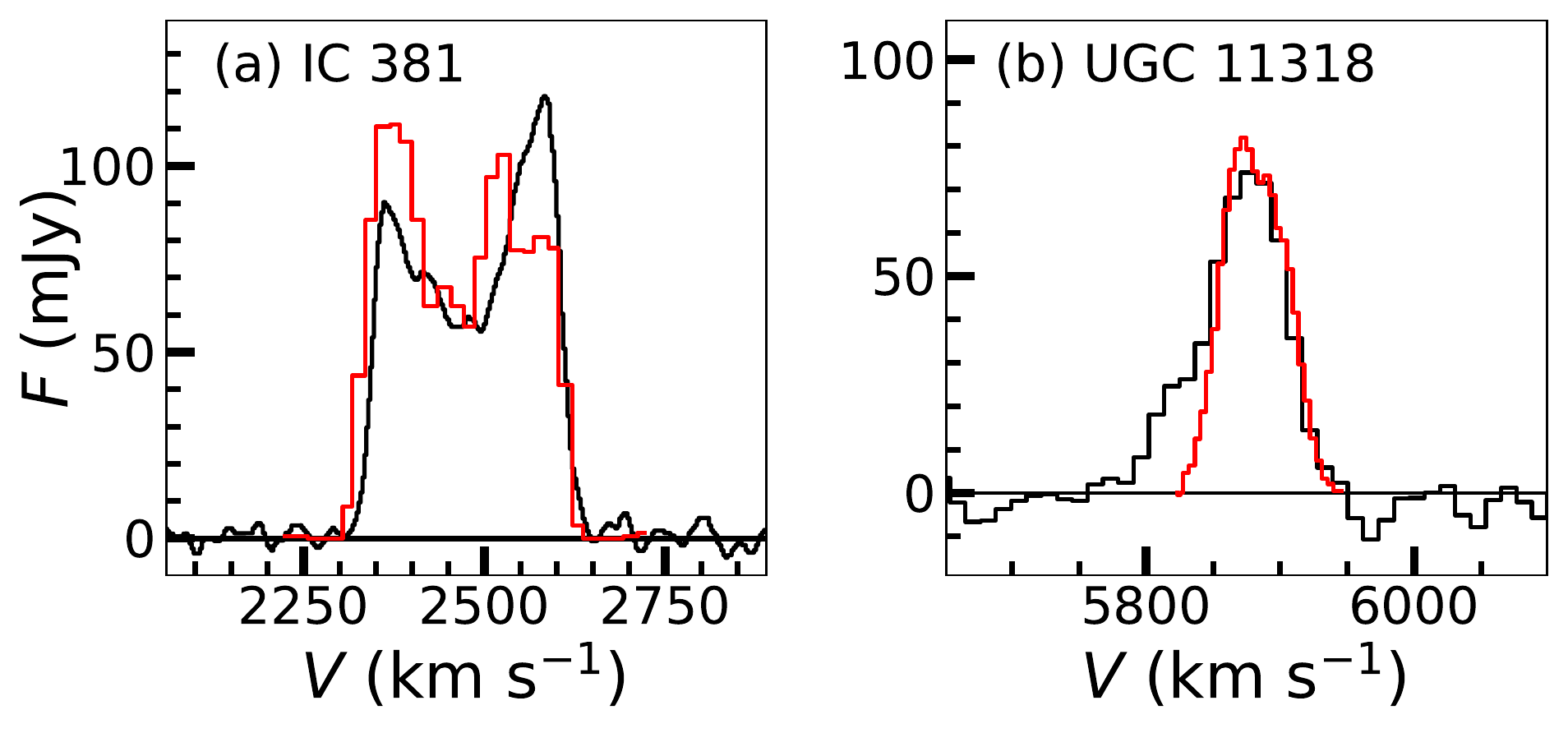}}
\caption{Comparison of the global profiles of single-dish (black) and 
interferometric (red) observation.}
\label{fig:ctm}
\end{figure*}

Figure~\ref{fig:ctm} shows the single-dish and interferometric spectra of the two galaxies that lie beyond 3~$\epsilon$ in the relation between \vrot\ and $V_{85c}$ (Figure~\ref{fig:outliers}).

\end{document}